\documentclass[desactivate]{aa}

\usepackage{graphicx}
\usepackage{txfonts}
\usepackage{pdflscape}
\usepackage{soul}
\usepackage{makecell}
\usepackage{textgreek} 

\usepackage{tipa}
\usepackage[colorlinks=true,urlcolor=blue,citecolor=blue,linkcolor=blue]{hyperref}

\graphicspath{{./}{figures/}}

\begin{document} 

   \title{SpectroTranslator: a deep-neural network algorithm to homogenize spectroscopic parameters}

   \author{G. F. Thomas\inst{1,2}\orcid{0000-0002-2468-5521}, G. Battaglia\inst{1,2}\orcid{0000-0002-6551-4294}, F. Gran\inst{3}\orcid{0000-0002-9252-6899}, E. Fern\'andez-Alvar\inst{2,1}, M. Tsantaki (M. T\textsigma\textalpha\textnu\texttau\textipa{\'\textalpha}\textkappa\texteta)\inst{4}\orcid{0000-0002-0552-2313}, E. Pancino\inst{4,5}, V. Hill\inst{3}, G. Kordopatis\inst{3}\orcid{0000-0002-9035-3920}, C. Gallart\inst{1,2}\orcid{0000-0001-6728-806X}, A. Turchi\inst{4}\orcid{0000-0003-3439-8005},\and T. Masseron\inst{1,2}}

   \institute{Instituto de Astrof\'isica de Canarias, E-38205 La Laguna, Tenerife, Spain \and Universidad de La Laguna, Dpto. Astrof\'isica, E-38206 La Laguna, Tenerife, Spain
   \and Universit\'e C\^ote d'Azur, Observatoire de la C\^ote d'Azur, CNRS, Laboratoire Lagrange, Bd de l'Observatoire, CS 34229, 06304 Nice cedex 4, France
   \and INAF – Osservatorio Astrofisico di Arcetri, Largo Enrico Fermi 5, 50125 Firenze, Italy
   \and Space Science Data Centre – ASI, Via del Politecnico SNC, 00133 Roma, Italy\\
              \email{gthomas@iac.es}
             }

   \date{\today}

\authorrunning{Thomas G. et al.}
  \abstract
   {The emergence of large spectroscopic surveys using different instruments and dedicated pipelines requires homogenising on the same scale the quantities measured by these surveys in order to increase their scientific legacy.}{We developed the {\sc SpectroTranslator}, a data-driven deep neural network algorithm that can convert spectroscopic parameters from the base of one survey (base A) to that of another one (base B).}{The {\sc SpectroTranslator} is constituted of two neural networks: an {\it intrinsic} network where all the parameters play a role in computing the transformation, and an {\it extrinsic} network, where the outcome for one of the parameters depends on all the others, but not the reverse. The algorithm also includes a method to estimate the importance that the various parameters play in the conversion from base A to B.}{As a showcase, we apply the algorithm to transform effective temperature, surface gravity, metallicity, [Mg/Fe] and line-of-sight velocity from the base of GALAH DR3 into the APOGEE-~2 DR17 base. We demonstrate the efficiency of the {\sc SpectroTranslator} algorithm to translate the spectroscopic parameters from one base to another using parameters directly by the survey teams, and are able to achieve a similar performance than previous works that have performed a similar type of conversion but using the full spectrum rather than the spectroscopic parameters, allowing to reduce the computational time, and to use the output of pipelines optimized for each survey. By combining the transformed GALAH catalogue with the APOGEE-~2 catalogue, we study the distribution of [Fe/H] and [Mg/Fe] across the Galaxy, and we find that the median distribution of both quantities present a vertical asymmetry at large radii. We attribute it to the recent perturbations generated by the passage of a dwarf galaxy across the disc or by the infall of the Large Magellanic Cloud.}{Although several aspects still need to be refined, in particular how to deal in an optimal manner with regions of the parameter space meagrely populated by stars in the training sample, the {\sc SpectroTranslator} already shows its capability and promises to play a crucial role in standardizing various spectroscopic surveys onto a unified basis.}

   \maketitle

\section{Introduction}

Precise physical quantities such as line-of-sight velocity, stellar atmospheric parameters and detailed chemical composition \footnote{Hereafter, we will refer to these quantities with "spectral parameters".}, derived from spectroscopic observations of individual stars, play a crucial role in the field of Galactic Archaeology by allowing us to constrain the mechanisms that drive the formation and the evolution of the Milky Way and of its neighbours. To date, a plethora of stars have already been observed by large spectroscopic surveys, either at low/medium resolution, such as the Sloan Extension for Galactic Understanding and Exploration \citep[SEGUE;][]{yanny_2009}, the Large sky Area Multi Object fiber Spectroscopic Telescope \citep[LAMOST;][]{zhao_2012a,yan_2022}, the RAdial Velocity Experiment \citep[RAVE;][]{steinmetz_2006}, Gaia-RVS \citep{recio-blanco_2023},  and DESI \citep{flaugher_2014, cooper_2023}, or high-resolution such as the Apache Point Observatory Galactic Evolution Experiment \citep[APOGEE;][]{abdurrouf_2022}, the Galactic Archaeology with HERMES \citep[GALAH;][]{buder_2021}, and the Gaia-ESO survey \citep{gilmore_2022,randich_2022}. This number is going to drastically increase in the coming years with the new generation of large spectroscopic surveys such as the WHT Enhanced Area Velocity Explore \citep[WEAVE;][]{dalton_2012,jin_2023}, the 4-metre Multi-Object Spectrograph Telescope survey \citep[4-MOST;][]{dejong_2019}, and the Sloan Digital Sky Survey-V \citep[SDSS-V;][]{kollmeier_2017}, which are going to observe millions of stars at high and medium resolution in both hemispheres, providing a more complete coverage of our Galaxy, and a necessary complement to the Gaia mission.

Each of these surveys employ different instruments, of varying wavelength coverage and spectral resolving power, and rely on their own dedicated data reduction and spectral analysis pipelines to yield spectroscopic parameters. Consequently, while many stars may overlap across these surveys, significant systematic differences exist in the derived spectroscopic parameters \citep[e.g.][]{hegedus_2023}. Even when considering identical data, the use of different pipelines can result into very dissimilar values of spectroscopic parameters due to variations in the methodology for spectral analysis or of the grids of synthetic spectra used \citep{allendeprieto_2016}. These discrepancies are not trivial and can potentially lead to misinterpretations of the observed chemical patterns when parameters derived from different surveys are used jointly \citep[see][and references within for a review on this problem]{jofre_2019}. This is particularly problematic given that several surveys possess different sky coverage and sample distinct volumes of our Galaxy, and are {\it de facto} complementing one another. Therefore, it is crucial to standardize spectroscopic parameters across surveys onto a unified scale.

Efforts have been made in recent years to develop generic data-driven methods capable of deriving spectroscopic parameters from spectra obtained by different surveys, enabling (at least partial) calibration onto a common scale. Examples include the {\sc Cannon} \citep{ness_2015,casey_2016,ho_2017}, the \textsc{Payne} \citep{ting_2019}, and \textsc{StarNet} \citep{fabbro_2018,bialek_2020}. For instance, \citet{wheeler_2020} used the {\sc Cannon}  to combine LAMOST and GALAH on the same scale; \citet{nandakumar_2022} used the same method to combine APOGEE and GALAH and \citet{xiang_2019} used the \textsc{Payne} to combine LAMOST with APOGEE and GALAH. \citet{guiglion_2024} recently employed a convolutional neural network to derive stellar parameters and individual abundances from the Gaia XP coefficients combined with the public Gaia RVS spectra expressed in the APOGEE base. These methods typically operate on low-level products in terms of processing (such as raw spectra, continuum-subtracted spectra, or the XP coefficients for Gaia) and this requires dealing with large amount of data and/or a heavy computational load. However, \citet{tsantaki_2022} presented a method operating on one of the end products of the spectroscopic pipelines, i.e. radial velocity measurements, and generated a homogeneous catalogue from six large spectroscopic surveys.

In this paper, we introduce the \textsc{SpectroTranslator}, a publicly available data-driven deep neural network algorithm designed to work on high-level spectroscopic information, to translate such parameters from the base of one survey to that of another survey. In Section~\ref{sec:presentation}, we present the \textsc{SpectroTranslator} algorithm, its architecture, and its range of applications. In Section~\ref{sec:testcase}, we present an application of the \textsc{SpectroTranslator} algorithm by transforming the effective temperature (T$_\mathrm{eff}$), surface gravity ($\log(g)$), metallicity ([Fe/H]), and magnesium abundance ([Mg/Fe]) from the GALAH DR3 survey into the base of APOGEE-~2 DR17. The importance of each parameter in the transformation is estimated in Section~\ref{sec:shap} using a method initially developed for multi-player cooperative games. We carry out astrophysical validation of the performance of the \textsc{SpectroTranslator} algorithm using globular clusters in Section~\ref{sec:GCs}. In Section~\ref{sec:sciencecase}, we present an example of the science enabled by such homogenization of spectroscopic catalogues, combining the APOGEE-~2 and the transformed GALAH samples to probe the distribution of [Fe/H] and [Mg/Fe] across the Milky Way. Finally, our conclusions are given in Section~\ref{sec:conclusions}.

\section{The {\sc SpectroTranslator} algorithm} \label{sec:presentation}

\begin{figure}
\centering
  \includegraphics[angle=0, viewport= 0 0 645 765,clip,width=8cm]{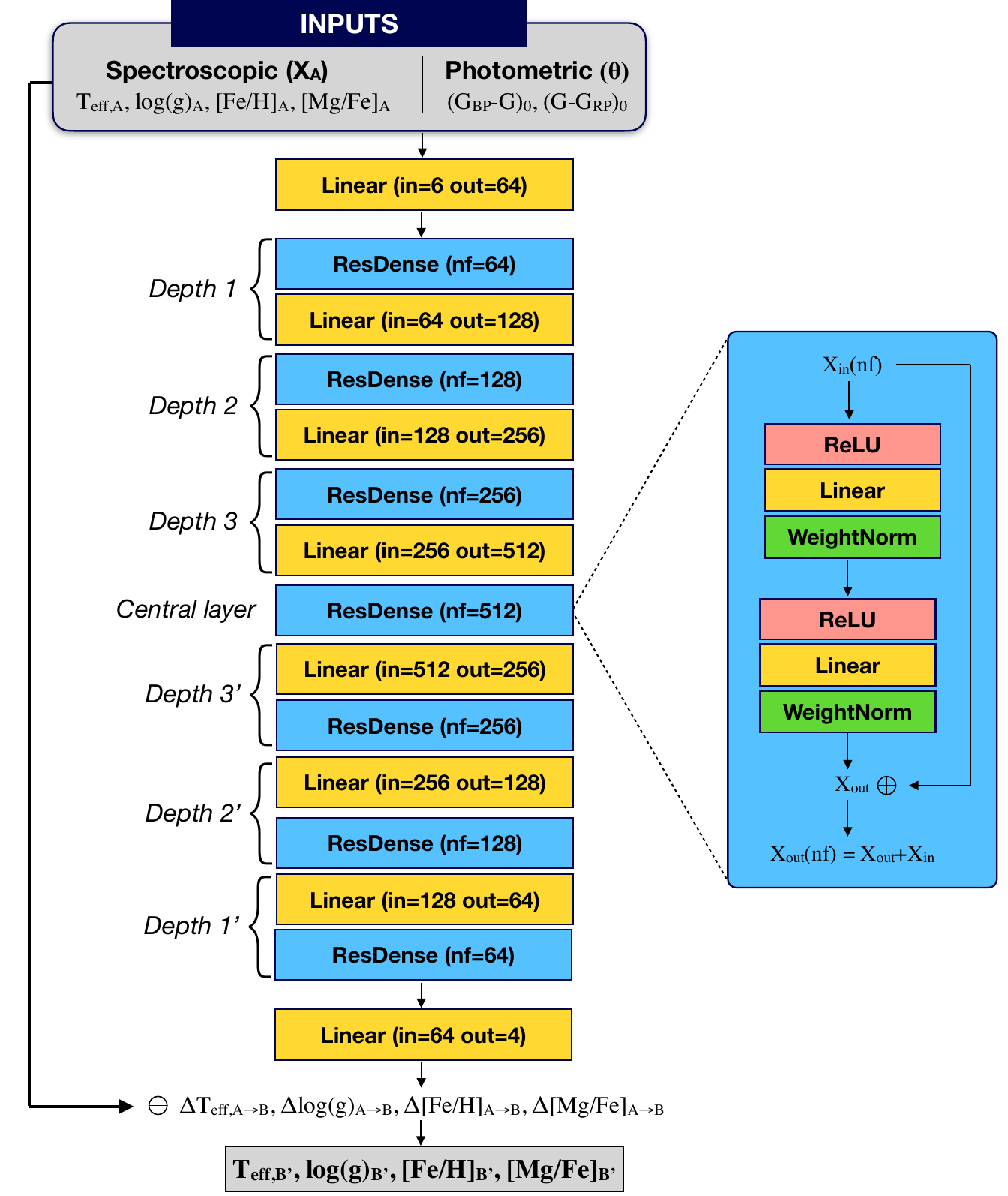}
   \caption{Sketch of the main part of the {\sc SpectroTranslator} algorithm. The difference between the spectroscopic parameters of catalogue {\it A} and {\it B} ($\bf{\Delta X_{A\rightarrow B}}$) are computed in a deep fully connected network, where the inputs are composed of the spectroscopic parameters from catalogue {\it A}, and of photometric colours. The deep network of depth-$n$ is made by a series of $n$ blocks of progressively increasing complexity, until a central layer of maximum complexity, before decreasing symmetrically. Each block is composed of a {\sc ResNet}-like unit (see Sect.~\ref{sec:archi}), illustrated on the right part, and of a linear layer that adjust the number of feature ($nf$) between each depth. Finally, the difference predicted by the deep network are added to the spectroscopic parameters of catalogue {\it A} to compute the parameters values expressed in the base of catalogue {\it B}.}
\label{fig:sketch}
\end{figure}

The goal of the {\sc SpectroTranslator} algorithm is to ``translate'' the values of spectroscopic parameters (for example, $T_\mathrm{eff}$, log$(g)$, line-of-sight velocity) of stars in catalogue $A$ expressed in the base {\it A}, $\\mathbf{X_A}$, into the base of another catalogue {\it B}, 
${\mathbf{X_B}}'$
; or said differently, to homogenise the values of the spectroscopic parameters of both catalogues $A$ and $B$ on the same base (here on the base of catalogue $B$). One could make the analogy of translating a sentence written in language {\it A} to language {\it B}, with the major difference in our case being that each word of the sentence (the spectroscopic parameters) are ordered in the same way and have the same function in both languages. 

This ``translation'' of the spectroscopic parameters from base {\it A} to base {\it B} can simply be expressed as:
\begin{equation}
{\mathbf{X_B}}' = \bf{X_A} + \bf{\Delta X_{A\rightarrow B}} \ ,   
\end{equation}
 $\bf{\Delta X_{A\rightarrow B}}$ is the value that we aim to determine, and is equal to the difference of each parameter between both bases. This value is derived using stars common to both Catalogue A and Catalogue B and by ensuring that the spectroscopic parameters translated from base A into base B ($ {\mathbf{X_B}}'$) closely align with the original values of those parameters listed in catalogue B ($\bf{X_B}$) for these stars.

 The core of the {\sc SpectroTranslator} algorithm is composed of two independent deep neural networks with the same architecture (described later in the section), trained using the stars in common between both catalogues. One of the networks is dedicated to transform the {\it intrinsic} parameters of a star, $\bf{X_{A, intr}}$, which in our case are the effective temperature, surface gravity, metallicity and chemical abundance (that is ${\bf X_{A, intr}} = [\mathrm{T_{eff,A}, \log(g)_A, [Fe/H]_A, [X/Fe]_A]}$), while the other network is dedicated to translate {\it extrinsic} parameters, $\bf{X_{A, extr}}$ (which in our case are only the line-of-sight, los, velocity; hence, ${\bf X_{A, extr}} = V_{los}$). The reason behind the choice of making two independent networks is that there is {\it a priori} no reason for the transformation of the {\it intrinsic} parameters from base $A$ to $B$ to be affected by the los velocity, while on the contrary, the transformation of the los velocity from base $A$ to $B$ can be affected by the {\it intrinsic} parameters, as shown recently by the Survey-Of-Survey team \citep[SoS,][]{tsantaki_2022}. However, it is interesting to note here that the {\it extrinsic} network can also be used to train other parameters, such as e.g. individual abundances, since the atmospheric parameters and the metallicity might impact the base transformation of individual elements, but likely the abundance of individual elements do not impact the transformation of these parameters. 

 Therefore, to make it explicit, we have:
 \begin{equation}
 \bf{X_{B,intr}'} = \bf{X_{A,intr}} + \bf{\Delta X_{A\rightarrow B,intr}} \ ,   
\end{equation}
and
\begin{equation}
 \bf{X_{B,extr}'} = \bf{X_{A,extr}} + \bf{\Delta X_{A\rightarrow B,extr}} \ ,   
\end{equation}
where ${\bf X_{A, intr}} = [\mathrm{T_{eff,A}, \log(g)_A, [Fe/H]_A, [X/Fe]_A]}$ and ${\bf X_{A, extr}} = \mathrm{V_{los,A}}$.

For each translated parameter $i$, we define $\mathbf{\Delta X_{A\rightarrow B, i}} = \mathrm{f}(\mathrm{X_{A,i}}, ..., \mathrm{X_{A,n}},\, $\boldmath$\theta$\unboldmath$)$, where \boldmath$\theta\ $\unboldmath are the features that are used to determine the transformation from base A to B. Therefore, we allow for each element of the vector $\bf{\Delta X_{A\rightarrow B}}$ to be dependent on the parameters in $\bf{X_A}$ and additional information contained in \boldmath$\theta\ $\unboldmath (like e.g. photometric colours), and not necessary linearly. This is motivated, for example, by the recent work of \citet{tsantaki_2022}, who show that the difference between the l.o.s velocity measured by different surveys and the Gaia measurement depends linearly on the metallicity, but quadratically on the effective temperature and on the magnitude G. 
 For what concerns the {\it intrinsic} network,  $\bf{\Delta X_{A\rightarrow B, intr}}= \mathrm{f}(\mathbf{X_{A,intr}}, $\boldmath$\theta_\mathrm{intr}$\unboldmath$)$, with \boldmath$\theta_\mathrm{intr}\ $\unboldmath containing the photometric colours, here  $\mathrm{(BP-G)_0}$ and $\mathrm{(G-RP)_0}$. For the {\it extrinsic} network, $\bf{\Delta X_{A\rightarrow B, extr}}= \mathrm{f}(\bf{X_{A,extr}}, $\boldmath$\theta_\mathrm{extr}$\unboldmath$)$, with \boldmath$\theta_\mathrm{extr}\ $\unboldmath containing $[\mathrm{T_{eff,A}, \log(g)_A, [Fe/H]_A, [X/Fe]_A]}$ and the photometric colours. 

 \subsection{Architecture of the network} \label{sec:archi}

 Instead of using a more classical multilayer perceptron network, we privilege a Residual neural Network \citep[{\sc ResNet},][]{He_2015}, to compute $\bf{\Delta X_{A\rightarrow B}}$. This is motivated by the fact that the latter are usually more stable and more robust than the former by limiting the {\it gradient vanishing} problem \citep{hochreiter_2003}, allowing to make deeper, and so more complex, models. In a {\sc ResNet}, the non-linear part of the network computes only the differences (residuals) between the initial values of the parameters and the values after the transformation, since these residuals are then added to the initial values via a {\it shortcut} that connects them to (some of) the inputs. As this architecture and the principles behind it are the same as the objectives of our algorithm (i.e. to compute the transformation from base $A$ to $B$), this also motivated our choice of using this type of network.

The architecture of the network used by the {\sc SpectroTranslator} algorithm, illustrated in Fig.~\ref{fig:sketch}, is strongly inspired by the {\sc ACTIONFINDER} algorithm presented in \citet{ibata_2021a}, but with some notable differences. Foremost among these is the fact that, unlike the {\sc ACTIONFINDER}, the {\sc SpectroTranslator} algorithm is in its entirety designed as a {\sc ResNet}-type network, with a {\it shortcut} connecting $\bf{X_A}$ to the residuals $\bf{\Delta X_{A\rightarrow B}}$; these are computed from $\bf{X_A}$ and \boldmath$\theta\ $\unboldmath through a fully connected network of depth-$n$\footnote{We decided to define the depth of the network ($n$) such that it corresponds to the number of blocks needed to reach the {\it central layer}, as the network is constructed symmetrically around it.}. The purpose of the initial layer is to take the input features and to increase the number of parameters in preparation of the next layer through a fully connected linear layer\footnote{A linear layer performs a linear transformation of the inputs ($x$) such that $y = W x + b$, where $W$ are the weights and $b$ the bias, and are learned by the algorithm. Therefore, they are equal to a fully connected dense layer without any activation function.}. Deeper layers are a succession of $n$-blocks constituted of a {\sc ResNet}-like unit and of a linear layer whose purpose is to connect each block together. The number of features used in each block incrementally increases by a factor 2, up to a {\it central layer} constituted of a unique {\sc ResNet}-like unit at depth $n+1$, where the number of features (nf) used is at is maximum and equal to nf$_1 \times2^{n}$ (512 in the example shown), where nf$_1$ is the number feature in the first hidden layer (64 in the example). After this central layer, the blocks decrease in size symmetrically, and a final linear layer takes the output of the last block and computes the residual of each spectroscopic parameter, $\bf{\Delta X_{A\rightarrow B}}$; they are then added to $\bf{X_{A}}$ to compute the transformed values expressed on the base of catalogue $B$ $(\bf{X_{B}})$.
Following the setup made by \citet{ibata_2021a}, each {\sc ResNet}-like unit is constituted of a Rectified Linear Unit \citep[ReLU][]{fukushima_1975,glorot_2011} activation function layer that feeds a fully connected linear layer followed by a weight normalization layer \citep{salimans_2016} that helps to improve the convergence of the network, all repeated twice. The sketch of a {\sc ResNet}-like unit is shown on the right side of Fig.~\ref{fig:sketch}. It has to be noted that to prevent overfitting, {\sc ResNet}-like units where nf$_i\leq 256$ include a {\it Dropout} layer \citep{hinton_2012,srivastava_2014} after the two ReLU layers to set randomly half of the weight to zeros.

Technically, the {\sc SpectroTranslator} is built using the {\sc Python} interface of the {\sc Keras} API \citep{chollet_2015} and the {\sc TensorFlow2} platform \citep{abadi_2016}. The algorithm has been built to be very flexible, in the choice of parameters that one wants to ``translate'', but also in its architecture to be able to adapt to the different needs that one can encounter.

\subsection{Range of application of the algorithm} \label{sec:quality}

Due to its conception, the {\sc SpectroTranslator} algorithm has to be trained on a subset of stars in common between two surveys, i.e. the training set. However, the parameter space covered by this training set is not necessary representative of the individual parameter space covered by these two surveys. This difference between them might lead to misinterpretation when the trained algorithm is applied to the entire catalogue of the input survey, as the coverage of its parameter space is likely larger than the range of parameter space covered by the sample used to train the algorithm. Therefore, it is crucial to estimate the range of parameters in which the transformations obtained by the trained algorithm are valid.   

For the {\sc SpectroTranslator} algorithm to be pertinent, it is essential to have a training sample sufficiently populated that represents well the parameter space of the union of the two surveys. This union sample does not necessarily need to be uniformly distributed across the parameter space, although this might require applying a weighting scheme to the data. However, it must adequately sample the full coverage of the parameter space.

It is noteworthy that the application domain of the {\sc SpectroTranslator} is independent of the fraction of the {\it output} survey (base B) covered by the training sample, provided the latter is sufficiently populated to encompass the full parameter space of the unions area between the input and output surveys. Indeed, as long as the latter point is correct, stars of the output survey (B) located outside the union area are beyond the scope of any stars from the input survey (base A), i.e. they do not have counterpart in the input survey, and so they are outside the domain of application of the {\sc SpectroTranslator} by default. It is essential to emphasize, however, that it is important to know the fraction of the {\it output} survey covered by the training sample if one intends to analyse data from the input survey (A) transformed by the {\sc SpectroTranslator} algorithm in the same way as for the output survey (B). Nevertheless, such analyses should be conducted on a case-by-case basis, and it is beyond the scope of this paper to present a universal method for performing this analysis.

On the other hand, it is crucial to know how representative is the training sample compared to the entire input catalogue (catalogue A), i.e. to know the domain of the parameter space where the algorithm is reliable to transform the data from catalogue $A$ to $B$. To estimate the domain of validity of the algorithm, we created a bit mask by binning the input parameter space. For the bins occupied by at least $N_\mathrm{thres}$ stars of the training sample, we can conclude that the transformation done by the algorithm is valid in the range covered by the bin, and their bitmask value is set to 1. For the bins that do not respect that condition, the results given by the algorithm are extrapolated, and they have to be treated carefully, as it is not possible to assert their validity in the range of parameters covered by the bin, and their bitmask values are set to 0. 

Binning the data in $n$-dimensions, where $n$ corresponds to the number of input parameters, can become quickly extremely costly in terms of computational memory, as the number of bins used for this task is proportional to $\mathcal{O}^n$. For this reason, the choice has been made to bin the parameter space for each possible combination of two input parameters, rather than binning the $n$ dimensional parameter space. This simplification comes at the cost of losing information on the correlation between more than two parameters, but has the advantage that the number of bins decreases to $\mathcal{O}(n)$. Thus, with this method, the validity of the transformation predicted by the algorithm for a given star is evaluated for all the possible combinations of two input parameters. In practice, for simplicity, the validity of the transformation is indicated by a boolean flag only if all the parameters are inside the parameter range of the training sample. However, the problematic set(s) of parameters are indicated in one of the columns ({\sc Qflag\_comments}, see Appendix~\ref{annex:metadata}) of the catalogue of transformed values.

\begin{figure*}
\centering
  \includegraphics[angle=0, viewport= 65 50 655 370,clip,width=12cm]{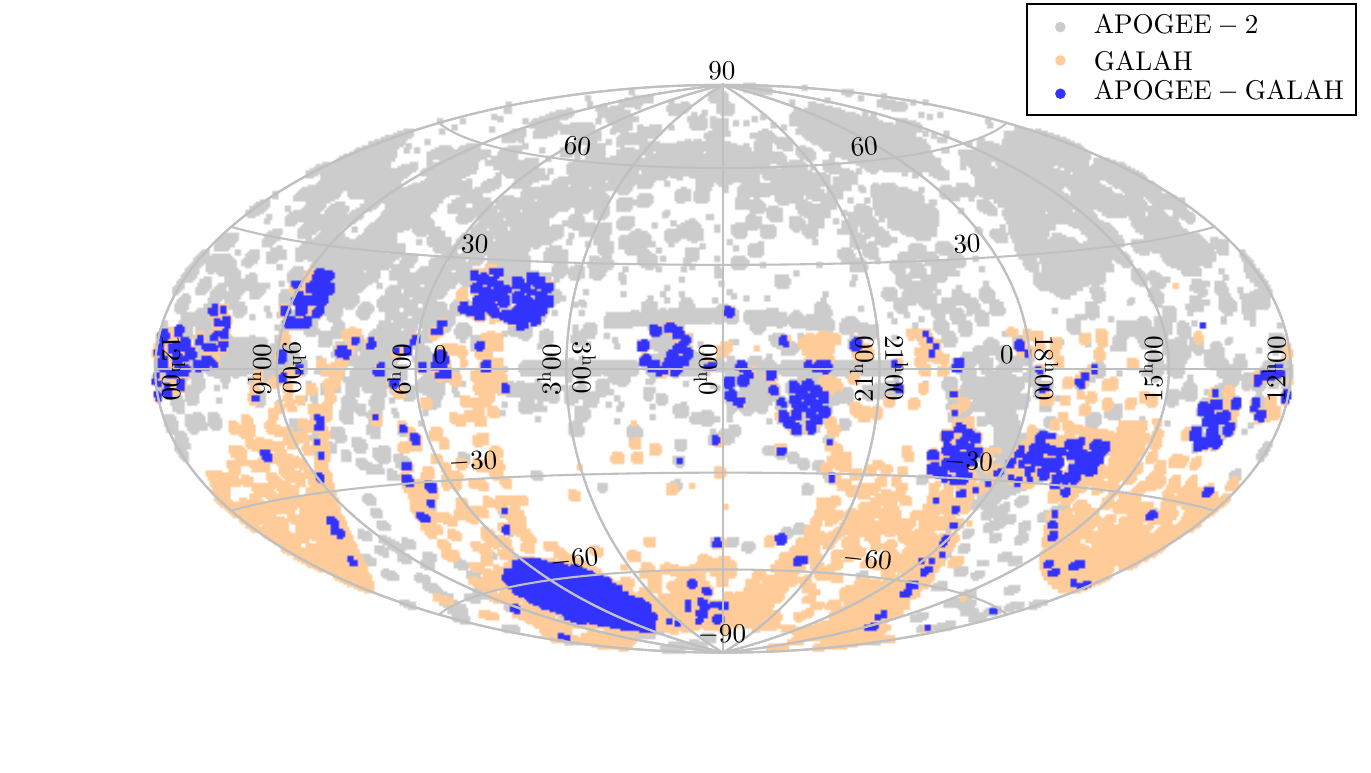}
   \caption{Sky coverage (in Galactic coordinates) of the $16,583$ stars in common (in blue) between APOGEE-~2 DR17 and GALAH DR3 that fullfil the criteria listed in Sect.~\ref{sec:data}, overlaid on the spatial coverage of these two surveys (in grey and orange, respectively).}
\label{fig:coverage}
\end{figure*}

In practice, in the {\sc SpectroTranslator} algorithm, the training sample defines the minimum and maximum range of validity of each parameter, as the algorithm is by definition defined between these ranges. Then, to make the different bitmasks, each parameter is decomposed in the same number of bins between these ranges. This number of bins has been set to 30 by default in the {\sc SpectroTranslator} algorithm after having tested different values using the training and validation sample of the test case presented in Section~\ref{sec:testcase}. However, a different value might be more suitable for different dataset, depending of the number of stars that they contain, the number of parameters or the range of these parameters. In the example shown in the next section, this led to a resolution of the application domain where the algorithm is valid of $\sim 110$ K for the effective temperature, $\sim 0.15$ dex for the surface gravity, $\sim 0.1$ dex in metallicity, $\sim 0.07$ dex for [Mg/Fe] and $\sim 0.05$ mag for the (BP$-$G) and (G$-$RP) colours. 

Following the same procedure, a series of bitmask are constructed from all the combinations of the output parameters. As said earlier, this is not made to estimate the representativity of the training sample in comparison to the entire output catalogue (catalogue B), but it helps to flag the stars located in regions of the parameter space where the {\sc SpectroTranslator} algorithm might not be fully reliable. This is particularly the case for the stars located near the border of the domain of application in the input parameter space.

\section{A test case: transforming GALAH to APOGEE} \label{sec:testcase}

In the following section, we present an example of application of the {\sc SpectroTranslator} algorithm to transform the effective temperature (T$_\mathrm{eff}$), surface gravity (log$(g)$), metallicity ([M/H]), magnesium abundance ([Mg/Fe]) and line-of-sight velocity (V$_\mathrm{los}$) from the GALAH catalogue (catalogue A) into the base of the APOGEE-~2 catalogue (catalogue B). Note that we have chosen to use the individual abundance ratio [Mg/Fe] instead of the global $\alpha$-abundance ratio ([$\alpha$/Fe]). This decision is based on the fact that the [$\alpha$/Fe] values obtained from GALAH and APOGEE-~2 may reflect the abundances of different elements due to their distinct wavelength ranges. In a comparison of values measured between APOGEE and optical measurements for a set of stars in common, \citet{jonsson_2018} found that magnesium exhibits the highest accuracy among $\alpha$-elements. Therefore, in this example application of the {\sc SpectroTranslator}, we preferred to use [Mg/Fe] rather than the global [$\alpha$/Fe] because it refers to the same chemical element between the two surveys and offers a more accurate scientific value for the science case presented in  Sect.~\ref{sec:sciencecase}. However, it is always possible to apply the {\sc SpectroTranslator} to transform the global $\alpha$-abundance ratio.

\subsection{Data} \label{sec:data}

\begin{figure}
\centering
  \includegraphics[angle=0,clip,width=7.5cm]{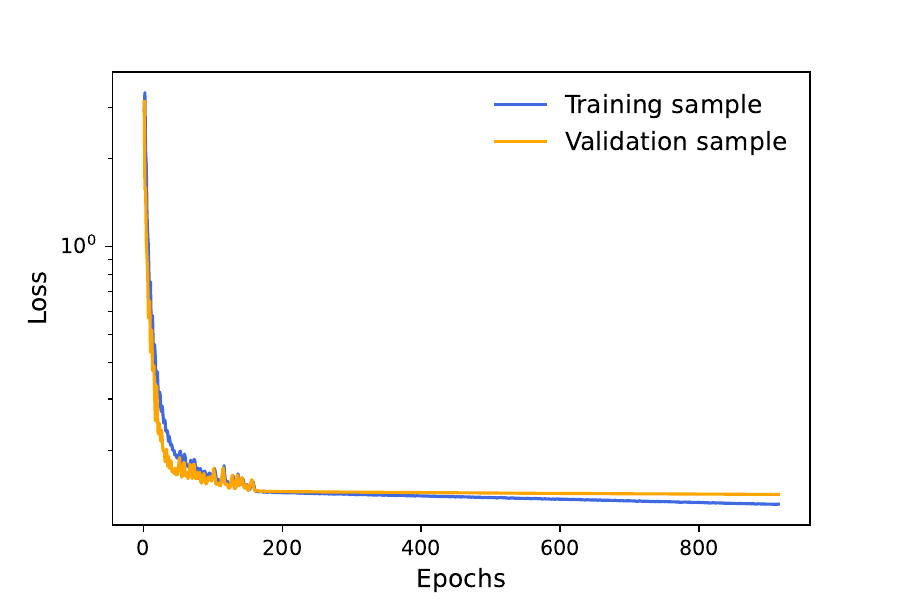}
   \caption{Evolution of the loss function as a function of the epoch for the intrinsic network. The loss of the training sample is shown by the blue curve, while the one of the validation sample is shown by the orange line.}
\label{fig:loss}
\end{figure}

\subsubsection{Catalogue A: GALAH DR3} \label{sec:GALAH}
The stellar parameters and chemical abundances on which we will apply the transformation are those from the main catalogue of the third data release of the GALAH survey \citep{buder_2021}. This catalogue contains  $588,571$ stars observed at high resolution ($R\sim 28,000$) in four non-continuous wavelength regions in the optical (between 4713-7887 \AA) using the High Efficiency and Resolution Multi-Element Spectrograph \citep[HERMES][]{sheinis_2015} mounted on the 3.9m Anglo-Australian Telescope (AAT). The data reduction pipeline used to derive the stellar parameters of the GALAH DR3 catalogue is mostly described in \citet{kos_2017}, with a few modifications listed in \citet{buder_2021}.

Following the best practices recommendations for GALAH DR3\footnote{\url{https://www.galah-survey.org/dr3/using_the_data/}}, and the work of \citet{hegedus_2023}, we selected only the stars respecting all the following criteria:
\begin{itemize}
    \item $S/N>30$ ({\sc snr\_c3\_iraf$>30$}),
    \item {\sc VBROAD} $<15~{\rm km}\,{\rm s}^{-1}$ to remove stars with significant rotation,
    \item have no flagged problems ({\sc flag\_sp=0}), 
    \item have valid [Fe/H] and [Mg/Fe] estimate ({\sc flag\_fe\_h=0} and {\sc flag\_mg\_fe=0}), 
    \item to have a corresponding entry in the Gaia DR3 catalogue, based on the crossmatch provided by the GALAH team.
\end{itemize}
The application of these criteria lead to a GALAH DR3 sample of 293,314 stars. The input spectroscopic parameters are {\sc teff, logg, fe\_h} and {\sc mg\_fe} for the {\it intrinsic} parameters, and {\sc rv\_galah} for the {\it extrinsic} network.

\subsubsection{Catalogue B: APOGEE-~2 DR 17} \label{sec:APOGEE}
The data used to define the base B are from the last data release (DR17) of the APOGEE-~2 \citep[Apache Point Observatory Galactic Evolution Experiment,][]{majewski_2017,abdurrouf_2022}. It contains $733,901$ stars observed at high resolution ($R\sim 22,500$) in near-infrared (15,140-16,940 \AA) by the APOGEE spectrographs \citep{wilson_2019} mounted on the Sloan 2.5m telescope of the Apache Point Observatory \citep{gunn_2006} and on the 2.5m Ir\'en\'ee du Pont telescope \citep{bowen_1973} at Las Campanas Observatory. The stellar parameters and chemical abundances have been obtained within the APOGEE Stellar Parameters and Chemical Abundances Pipeline \citep[{\sc ASPCAP}][]{garciaperez_2016}, from which we use {\sc teff}, {\sc logg}, {\sc m\_h} and {\sc mg\_fe} for the {\it intrinsic} network, and {\sc vhelio\_avg} as the output of the {\it extrinsic} network. 

As previously, following the recommendations for APOGEE DR17 \footnote{\url{https://www.sdss4.org/dr17/irspec}} and the work of \citet{hegedus_2023}, we selected only the stars respecting all the following criteria:
\begin{itemize}
    \item {\sc SNR>100}
    \item are not flagged as {\sc STAR\_BAD}\footnote{The {\sc STAR\_BAD} flag is set for a star if any of {\sc TEFF\_BAD}, {\sc LOGG\_BAD}, {\sc CHI2\_BAD}, {\sc COLORTE\_BAD}, {\sc ROTATION\_BAD}, {\sc SN\_BAD} or {\sc GRIDEDGE\_BAD occur} is set.}, {\sc FE\_H\_BAD}, and {ALPHA\_FE\_BAD} in the {\sc APOGEE\_ASPCAPFLAG} bitmask,
    \item the limitation of the scatter in los velocity {\sc vscatter} $<1~{\rm km}\,{\rm s}^{-1}$ in order to eliminate most binaries and other variable stars, 
    \item have a Gaia DR3 counter-part, with the cross-identification made by the APOGEE team.
\end{itemize}
This led to an APOGEE-~2 sample of $455,486$ stars. 

\begin{figure*}
\centering
  \includegraphics[angle=0,clip,width=8cm]{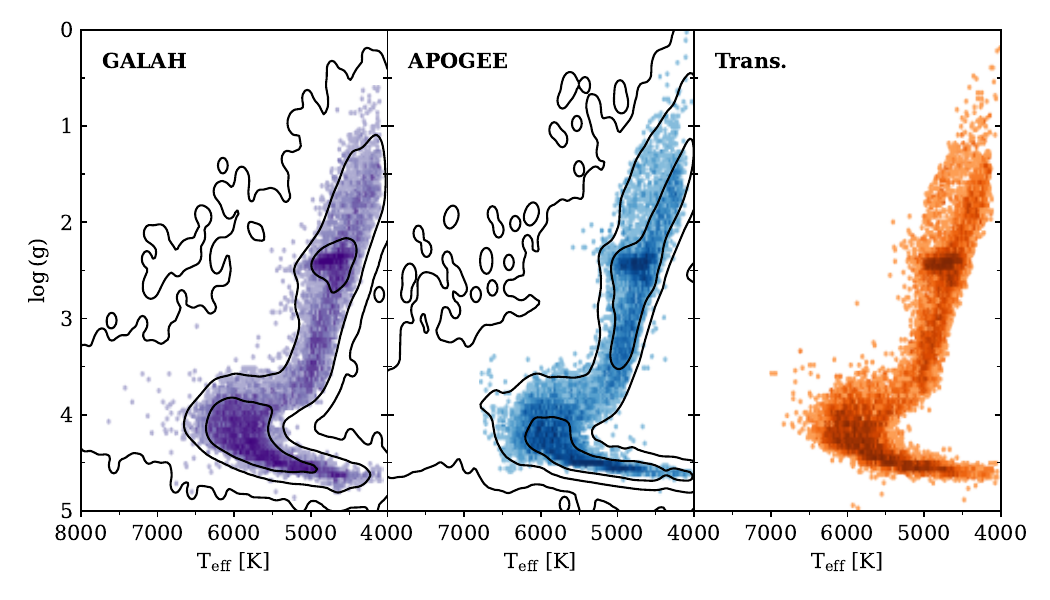}
  \includegraphics[angle=0,clip,width=8cm]{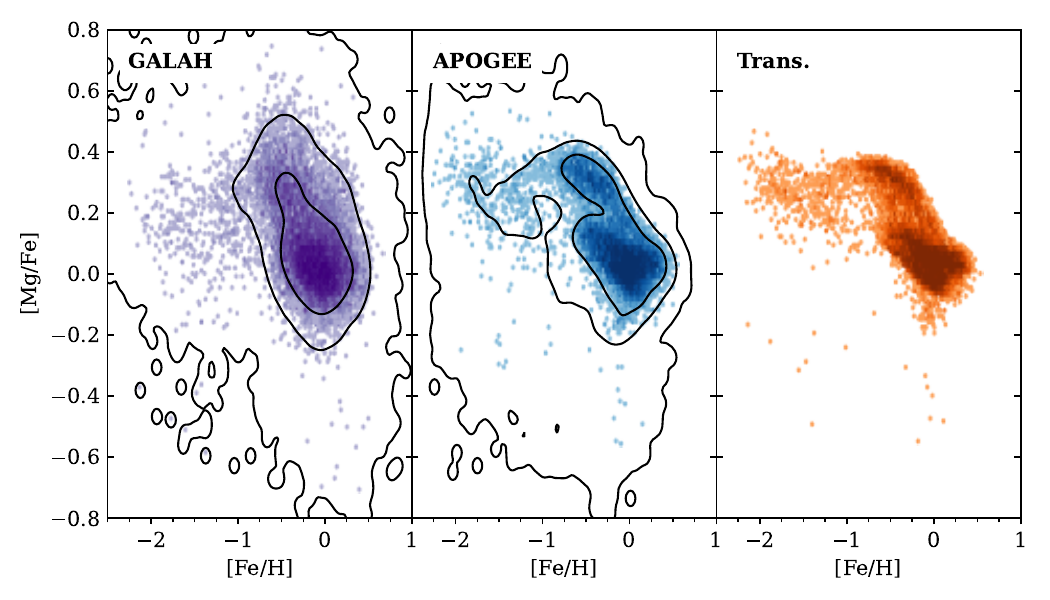}
  \includegraphics[angle=0,clip,width=8cm]{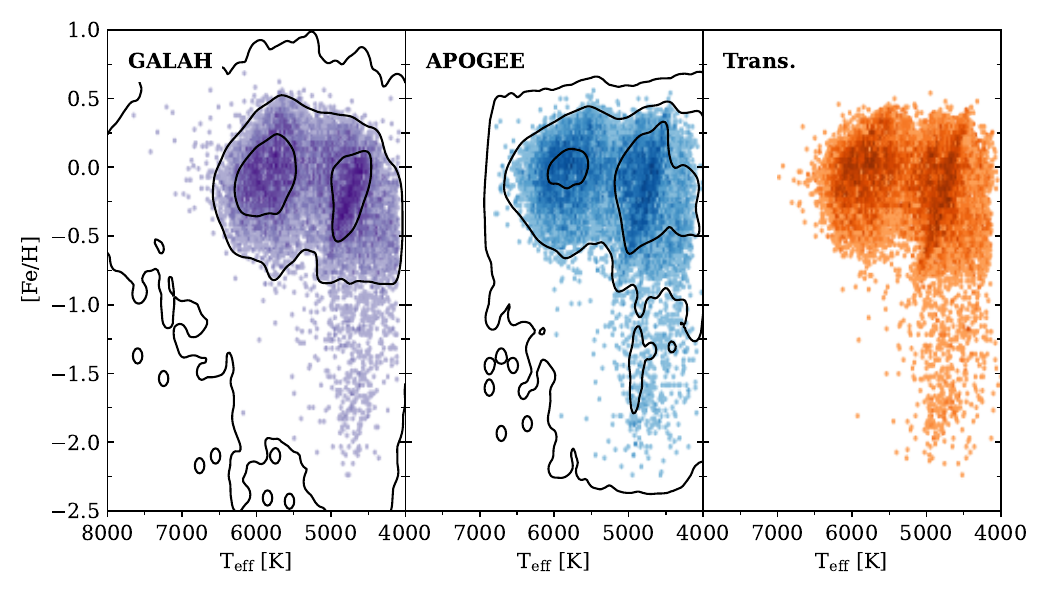}
  \includegraphics[angle=0,clip,width=8cm]{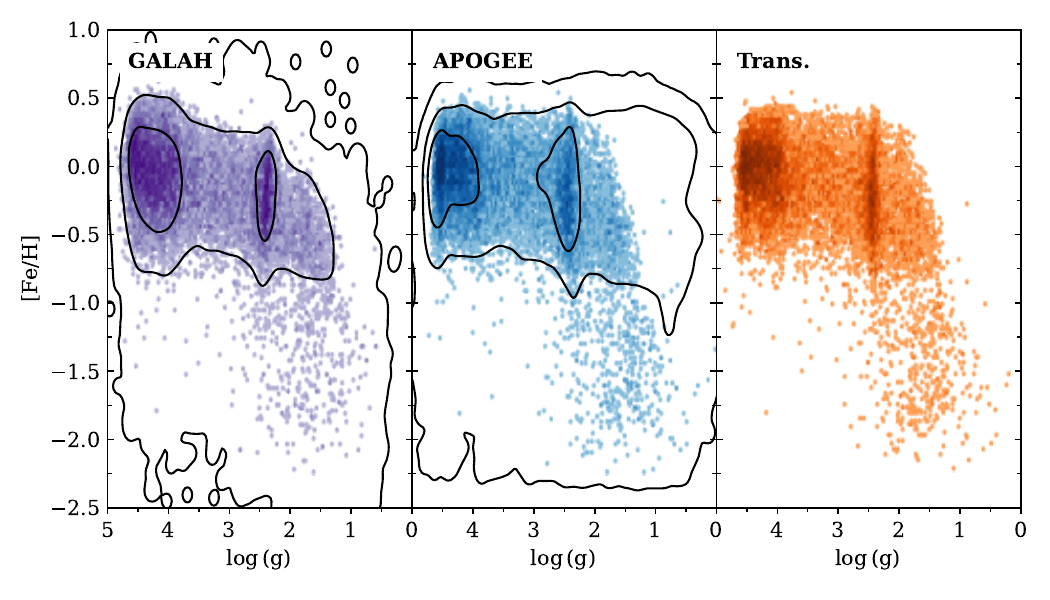}
  \includegraphics[angle=0,clip,width=8cm]{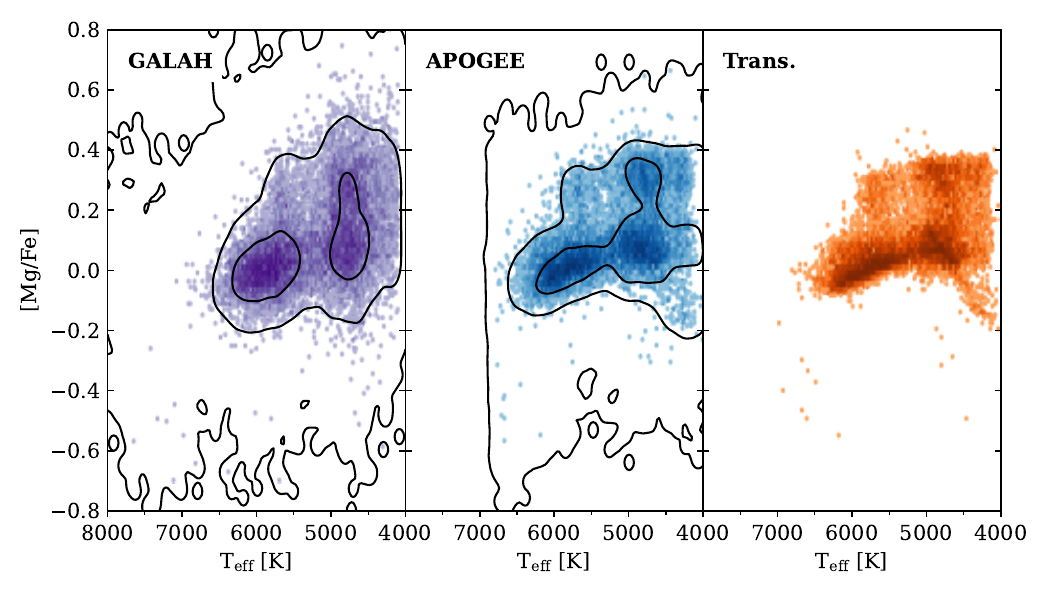}
  \includegraphics[angle=0,clip,width=8cm]{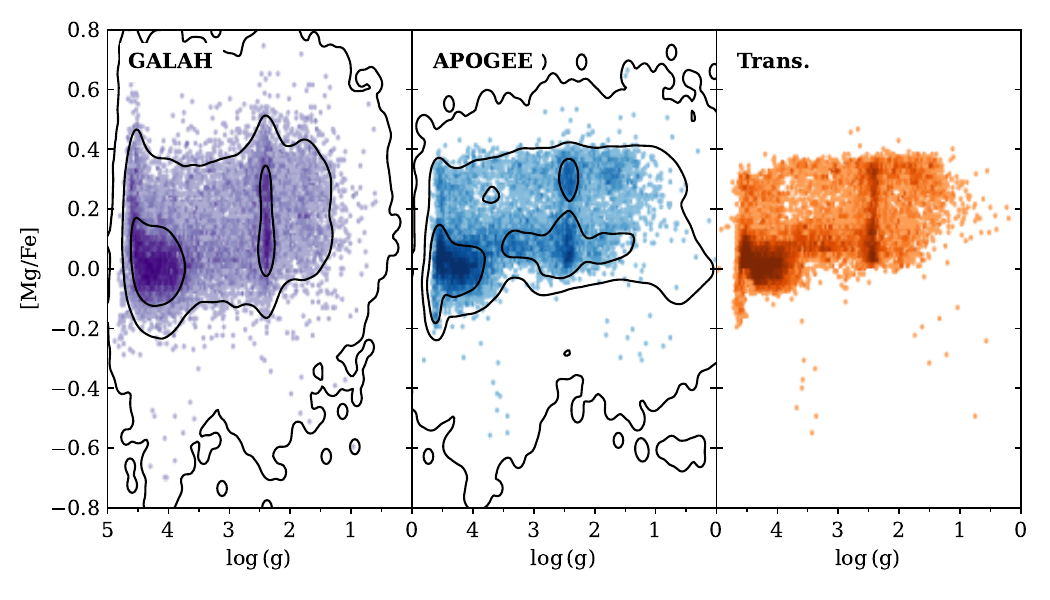}

   \caption{Comparison between the coverage of the parameter space for stars in the training sample using the original GALAH data (in purple, left panels), the original APOGEE-~2 data (in blue, middle panels) and the parameters transformed from the GALAH into the APOGEE-~2 base by the {\sc SpectroTranslator} algorithm (in orange, right panels). The contours in the right and middle panels depict the 1, 10, and 100 stars per bin limits in the selected GALAH DR3 catalogue (refer to Section~\ref{sec:GALAH}) and APOGEE DR17 (refer to Section~\ref{sec:APOGEE}), respectively. }
\label{fig:comparison}
\end{figure*}

\begin{figure*}
\centering
  \includegraphics[angle=0,clip,width=15cm]{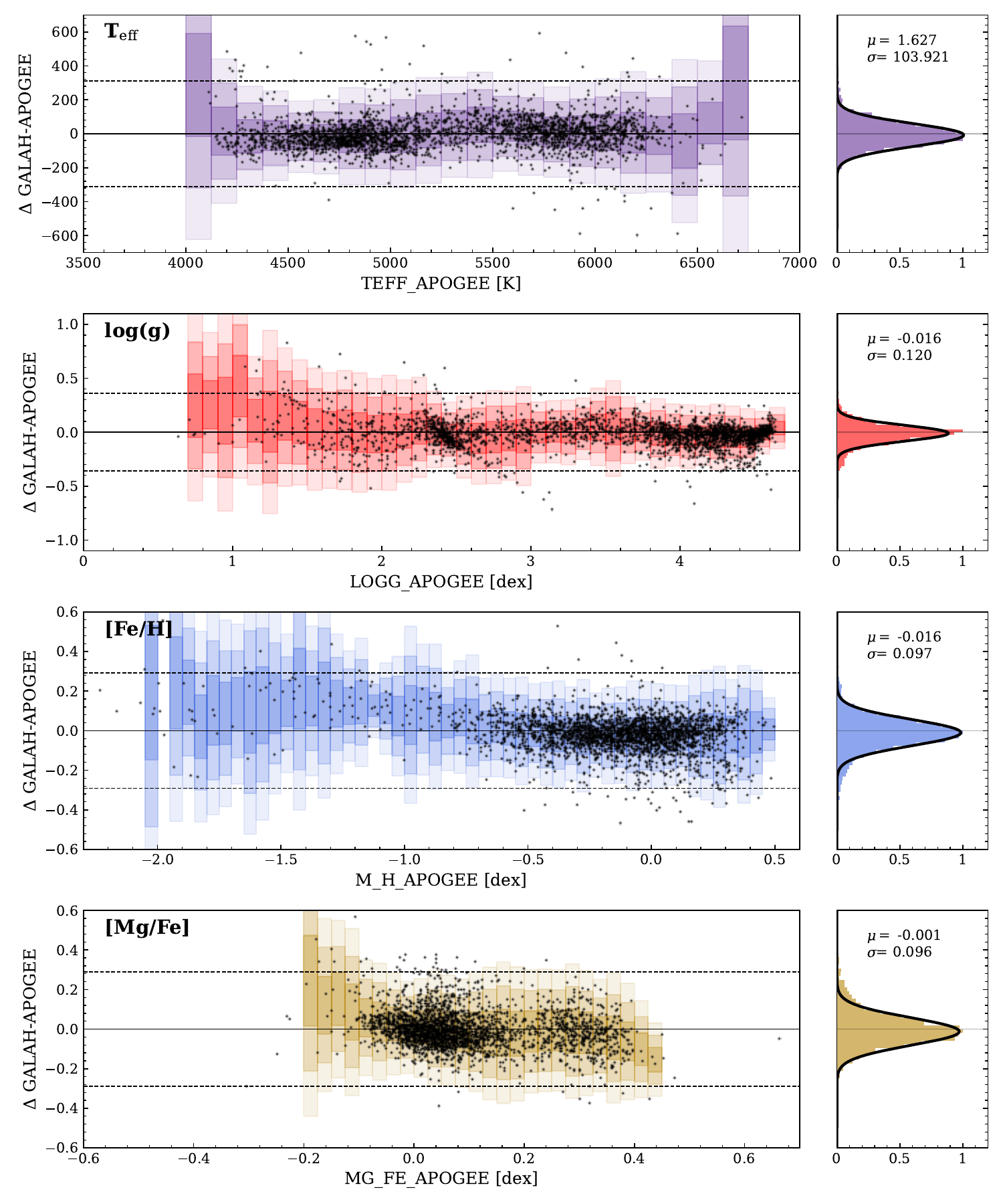}
  \caption{Variation of the residuals between ``original'' GALAH ($\bf{X_{A}}$) and APOGEE-~2 ($\bf{X_{B}}$) values for each output parameters. Left: The black points correspond to the validation set, while the shaded areas correspond to the 1, 2 and 3$\sigma$ around the local mean of the residual obtained from the training set. Bins where the number of stars of the training set is lower than 5 are not shown, due to the poor statistic in them. The horizontal dashed lines correspond to the average 3$\sigma$ of the residuals. Right panels: The coloured histograms show the difference between the ``original'' GALAH ($\bf{X_{A}}$) and APOGEE-~2 ($\bf{X_{B}}$), and black curve shows the Gaussian fitted to this histogram used to measure the average mean and standard deviation of the residuals, which are quoted on the right panels of each parameters.}
\label{fig:res_ori}
\end{figure*}

\subsection{The training setup} \label{sec:setup}
There are $16,583$ stars in common between the selected APOGEE-~2 and GALAH DR3 samples as selected in Sect.~\ref{sec:data}, which position is shown in Fig.~\ref{fig:coverage}. These stars constitute the basis for the training/validation sets. From these, we excluded all the stars located in regions of high extinction (E(B-V)$>0.3$) based on the extinction map of \citet{schlegel_1998} recalibrated by \citet{schlafly_2011}, as the Gaia colour in these regions -- mainly located close to the Galactic plane -- might be significantly affected by the extinction. Therefore, $13,664$ stars are used to train and validate the algorithm, separated between the training set, composed of $10,931$ stars (80\% of this initial sample) randomly selected, and of the validation sample composed of the other $2,733$  stars (20\%). Note here that the training set of the {\it intrinsic} network can be composed of different stars than the training set of the {\it extrinsic} network as these two networks are trained independently.

For the {\it intrinsic} network, the transformation from base A to B is computed from the spectroscopic parameters from GALAH $\bf{X_{A,intr}}=$[{\sc teff}, {\sc logg}, {\sc fe\_h}, {\sc mg\_fe}] and from the extinction corrected colours \boldmath$\theta_\mathrm{intr}$\unboldmath$=[$(BP-G)$_0$, (G-RP)$_0$], where the reddening conversion coefficients are adopted from \citet{marigo_2008}, following previous works \citep[e.g.][]{sestito_2019,thomas_2022}. The parameters expressed in base B (here APOGEE-2) are $\bf{X_B}=$[{\sc teff}, {\sc logg}, {\sc m\_h} and {\sc mg\_fe}] from APOGEE-~2. 

Note here that all the input and output parameters are normalized to have a distribution with a mean equal to zero and a standard deviation of one with respect to the training sample, done using the {\sc StandardScaler} of the {\sc Scikit-learn} library \citep{pedregosa_2011}. This process is standard when using neural networks, as it increases significantly the stability of the network, and avoids the loss function\footnote{The loss function is the metric that quantify the disparity between the transformed and the actual output values.} to be dominated by one of the parameters. The values and the parameterization of the {\sc SpectroTranslator} presented below have been optimised for the base transformation from GALAH to APOGEE-2. This parameterisation will be used by default for other transformations that we will provide in the future, unless specified otherwise. However, the design of the {\sc SpectroTranslator} is relatively flexible, allowing for adjustments in the number of layers used, the number of neurons per layer, or the loss function employed (see below), to accommodate transformations from other bases or for a different set of parameters to transform. As shown in Fig.~\ref{fig:sketch}, the adopted network has a depth of 3, and reaches a maximum of 512 neurones per layer at the central layer. The loss function used here is a mean absolute error (MAE, L1-regularization) function due to the potential high number of outliers, in particular at low metallicity where the number of stars is lower than at high metallicity; other loss functions like the mean square error (or L2-regularization) are more sensitive to outliers. The parameters (weights and biases) of each neurone that minimize the loss function are computed iteratively using the Adaptive Moment Estimation optimization method \citep[also known as Adam,][]{kingma_2014}, a modification of the classical stochastic gradient descent method that prevents falling into a local minimum. The algorithm is trained with three successive learning rate hyperparameters ($10^{-3}$, $10^{-4}$ and $10^{-5}$). To limit the amount of time needed to train the algorithm, the training phase is allowed to stop before it reaches the maximum of established epochs ($1,000$ here) if the loss value of the {\it validation} set did not improve during the last 20 epochs. In such case, the trained parameters used correspond to the parameters found at the epoch where the loss function of the validation set reached its minimum. The reason of choosing to monitor the loss function of the {\it validation} set instead of the one of the training set was made to prevent overfitting. With these conditions, in this specific case presented here, the {\it intrinsic} network needs to be typically trained over $\sim 900$ epochs in less than 30~minutes on a 8$\times$1.90 GHz machine, as one can see in Fig.~\ref{fig:loss}.

For the {\it extrinsic} network, a similar setup is adopted, with similar input as for the {\it intrinsic} network to which the los velocity of GALAH ({\sc rv}) has been added, and the output corresponds to the average l.o.s velocity of APOGEE-~2 ({\sc VHELIO\_AVG}). In addition, we removed the binary candidates listed in the catalogues of either \citet{price-whelan_2020} or \citet{traven_2020}, on top of removing stars with a scatter between different measurement in APOGEE-~2 of more than $1$~km~s.$^{-1}$. Despite these criteria, we found that some stars were having very high discrepancies between the l.o.s velocity measurement of GALAH and APOGEE-~2, which we attributed to potential binaries or other suspicious objects (e.g. pulsating stars). We therefore perform a 5-$\sigma$ clipping on the difference of measurement between GALAH and APOGEE-~2. This let us with a total of $12,266$ stars with, 9,813 (2,453) stars in the training (validation) sample of the {\it extrinsic} network.

\begin{figure*}
\centering
  \includegraphics[angle=0,clip,width=15cm]{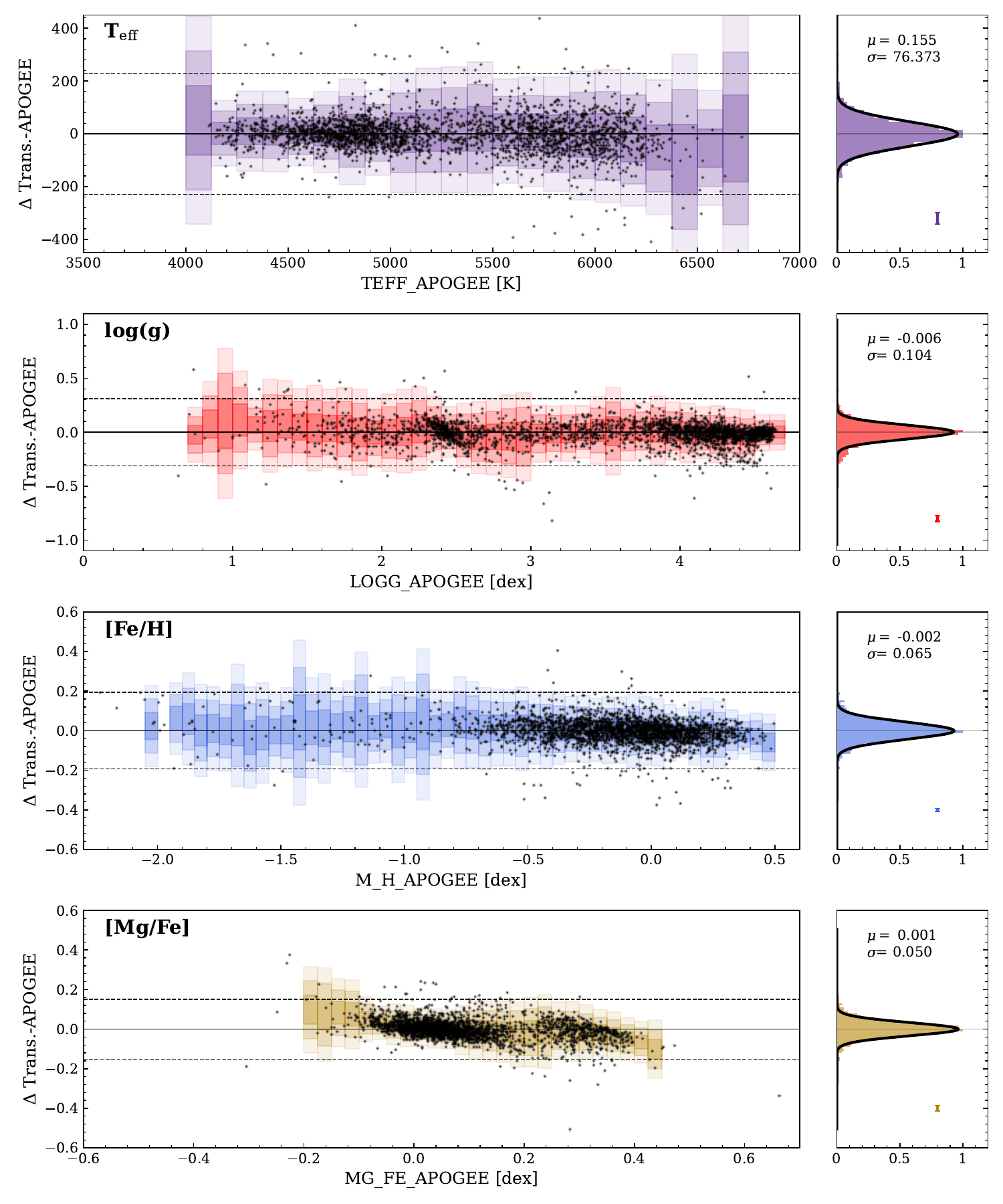}
   \caption{Similar to Fig.~\ref{fig:res_ori}, but with residuals between the transformed ($\bf{X_{B}'}$) and the ``original'' ($\bf{X_{B}}$) APOGEE-~2 values. For information, the average uncertainties of the APOGEE-~2 parameters in the training set is indicated by the error bar in the lower part of the right panels.}
\label{fig:residual}
\end{figure*}

\subsection{Results of the intrinsic network}

\subsubsection{Analysis} \label{sec:res_intrinsic}
Fig.~\ref{fig:comparison} presents a comparison of the parameter space covered by the training set using the spectroscopic values of the GALAH catalogue, ${\bf X_A}$  (in purple), of the ``original'' APOGEE-~2 catalogue ${\bf X_B}$ (in blue), and of the values transformed by the {\sc SpectroTranslator} algorithm, ${\bf X_B'}$ (in orange). A visual inspection of this figure shows that the coverage of parameter space for the spectroscopic parameters transformed by the algorithm is more similar to the coverage of the original APOGEE-~2 values than of that of the GALAH data, confirming the strength of the algorithm to correctly transform the data from one base to another. This is particularly visible on the [Fe/H]-[Mg/Fe] diagram where the thin/thick (low/high-$\alpha$) disc separation (around [Mg/Fe]$\sim 0.15$) is more enhanced with the transformed values than with the initial GALAH data, comparable to the separation visible with the APOGEE-~2 data. This enhancement of the thin/thick disc separation is also particularly visible in the $\log$(g)-[Mg/Fe] and T$_\mathrm{eff}-$[Mg/Fe] diagrams. In general, one can clearly see that the [Mg/Fe] parameter is the one which is the most affected by the change of base from GALAH to APOGEE-~2, and that this change seems to be correctly learned by the algorithm. However, there is also a significant difference in the metallicity between the GALAH and the APOGEE-~2 data, which seems to be correlated to the surface gravity, and in particular for the giant stars ($\log$(g)$<3.5)$, as they cover a wider range of metallicities in APOGEE-~2 than in the GALAH at a given $\log$(g). It is interesting to see that despite that difference between the two bases, the {\sc SpectroTranslator} algorithm is able to learn the transformation and to recover the wider spread seen in APOGEE-~2. The algorithm is also able to transform correctly the effective temperature and the surface gravity, as attested by the distribution of the red clump stars (around $\log$(g)$\simeq 2.5$) and of the metal-poor end of the top of the red giant branch (T$_\mathrm{eff}\sim 5000$ K and $\log$(g)$<2.3$).

This qualitative analysis of the performance of the {\sc SpectroTranslator} algorithm is confirmed quantitatively by comparing Fig.~\ref{fig:res_ori} and Fig.~\ref{fig:residual}. In Fig.~\ref{fig:res_ori}, we show the 
difference between the ``original'' GALAH ($\bf{X_{A}}$) and the APOGEE-~2 ($\bf{X_{B}}$) values for each parameter for the stars in the training/validation sample, while Fig.~\ref{fig:residual} shows the same relation but with the transformed (${\mathbf{X_B}}'$) instead of the ``original'' GALAH value. On these figures, the black points in the left panels correspond to the residuals for the stars of the {\it validation} set, while the residuals of the {\it training} set are illustrated by the shaded bins, corresponding to $1\sigma$, $2\sigma$, and $3\sigma$ around the local mean of the residual. This separated analysis of the residual for the training and validation sets is important, as we can see that the trained algorithm is not subject to overfitting since the trends visible in the variation of the residuals are similar between the training and validation sets for all the parameters. Note here that we do not show the trend of the training set for the region where the number of stars per bin is lower than 5 due to the poor statistical information that they hold. We can see that the GALAH parameters after the transformation into the APOGEE-2 base (${\mathbf{X_B}}'$) are well in agreement with the original APOGEE-2 values ($\bf{X_B}$), as for all the parameters, the mean of the residuals is around zero, with a scatter of $\sigma_{\mathrm{T_{eff}}}=76$ K, $\sigma_{\log\mathrm{(g)}}=0.104$ dex, $\sigma_\mathrm{[Fe/H]}=0.065$ dex and $\sigma_\mathrm{[Mg/Fe]}=0.050$ dex, significantly lower than the scatter of the difference between the ``original'' GALAH and APOGEE-~2 values $(\sigma_{\mathrm{T_{eff}}}= 103.9$ K, $\sigma_{\log\mathrm{(g)}}=0.120$ dex, $\sigma_\mathrm{[Fe/H]}=0.097$ dex, and $\sigma_\mathrm{[Mg/Fe]}=0.096$ dex $)$. Moreover, for the effective temperature, the surface gravity and the metallicity, the residual between the transformed and the ``original'' APOGEE-~2 values do no show any trend with the value of the corresponding parameter, while the difference between the ``original'' GALAH and APOGEE-~2 values present clear trends, in particular for the giant stars with low surface gravity and for the most metal-poor stars. For the Mg abundance, we can see that for [Mg/Fe]$<-0.15$~dex (in the APOGEE-~2 base), the transformed abundances are in average higher than in APOGEE-~2 by typically 0.1~dex, while for [Mg/Fe]$>0.4$~dex we observe the opposite trend. A similar trend, although twice more important, is present between the ``original'' GALAH and APOGEE-~2 values. The fact that the {\sc SpectroTranslator} is not able to completely remove this trend as it does with other parameters is likely the consequence of the low number of stars in the training sample present in these regions, 46 (0.42\%) and 47 (0.42\%) respectively. 

\begin{figure*}
\centering
  \includegraphics[angle=0,clip,width=16cm,viewport=80 0 935 430]{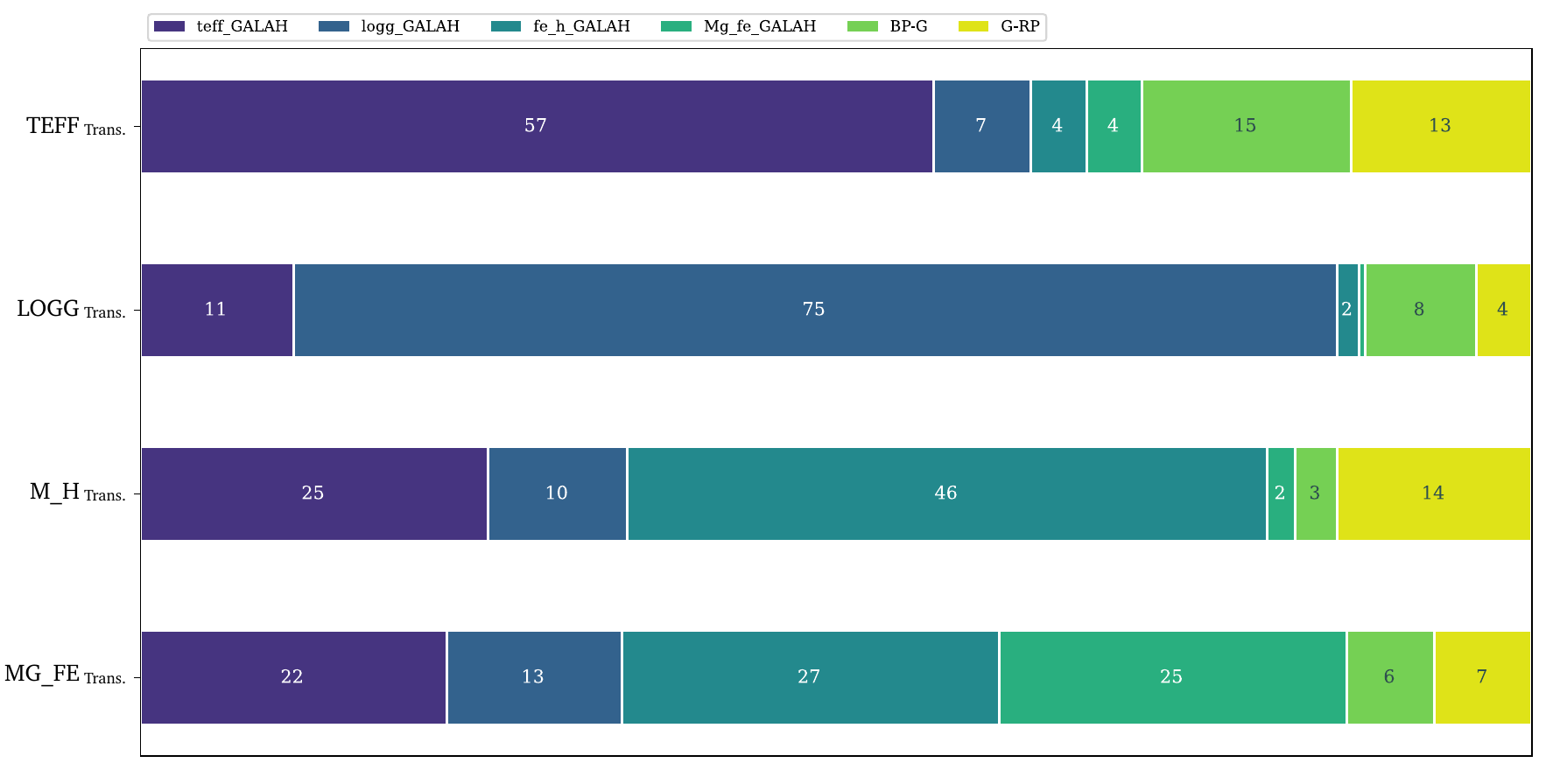}
   \caption{Relative importance that each feature has on the transformation of the parameters in the APOGEE-~2 base, computed from the mean absolute SHAP values (see Sect.~\ref{sec:shap}).}
\label{fig:shap_intrinsic}
\end{figure*}

The residuals that we obtained using the {\sc SpectroTranslator} are slightly larger than the residuals obtained by \citet{nandakumar_2022} who used directly the GALAH spectra and trained the {\sc Cannon-2} \citep{ness_2015,casey_2016} algorithm to obtain the spectroscopic parameters on the APOGEE base (refer to as GCAA in their paper), except for the surface gravity. Indeed, they found a residual between the values estimated by the {\sc Cannon} algorithm and the ``original'' APOGEE values of 58~K for the effective temperature, 0.12~dex for the surface gravity, 0.04~dex for the metallicity and 0.02 dex for the [$\alpha$/Fe] value. However, it is important to note here that they used the values from APOGEE-~2 DR16, and their [$\alpha$/Fe] parameter is the general $\alpha$-abundance, composed of a combination of several elements. This might explain the differences that we have regarding the [$\alpha$/Fe]-[Mg/Fe] residual.

\subsubsection{Feature importance} \label{sec:shap}

 In the previous section, we showed that the {\sc SpectroTranslator} algorithm is able to correctly learn the transformation from the GALAH to the APOGEE-~2 base, but we did not explore the contribution of each parameter in this transformation.
 
 Due to their complex non-linear nature, neural networks are in general hard to interpret, in particular regarding the importance that each input parameter (or feature) has. However, in the last years, partly motivated by the {\it right to explanations} established by the European Union \footnote{\url{https://eur-lex.europa.eu/eli/reg/2016/679/oj}}, a lot of progress has been made in the interpretability of these systems, and feature importance ranking has become an active research area \citep[e.g.][]{samek_2017,wojtas_2020}. Many methods are now available to interpret the importance that each input feature has in a deep neural network \citep[e.g.][]{tulioribeiro_2016,ribeiro_2018}. Among them, one of the most popular is the SHapley Additive exPlanations \citep[SHAP,][]{lundberg_2017} method, which is based on the optimization method developed by \citet{shapley_1953} to assign the payouts of each player in cooperative game theory, known as Shapley values. In general, the Shapley values are computationally expensive to measure, as they require retraining $2^n$ times the neural network, where $n$ is the number of input features. The SHAP estimation method encompass a range of different techniques, including the popular {\it LIME} method \citep{tulioribeiro_2016}, to approximate the Shapley values while reducing significantly the computational cost. The SHAP values indicate the contribution of each input feature (\boldmath$\mathrm{X_A},\theta$\unboldmath) to move the transformed values ($\mathrm{{\mathbf{X_B}}'}$) from the mean of the prediction.

 To compute the SHAP values and to estimate the importance (Imp) of each input feature, we used the {\sc KernelExplainer} method from the {\sc SHAP} python package\footnote{\url{https://shap.readthedocs.io/}.}. The ``typical'' input values of the trained {\sc SpectroTranslator} network are obtained by selecting randomly a subsample of 100 stars from the {\it training} set. These typical input values are then used to compute the individual SHAP values at the location in the parameter space of 100 stars randomly selected from the {\it validation sample} by performing 100 permutations. This results in a total of $100 \times 100 \times 100$ computations, and take a few minutes on a $8 \times 1.90$ GHz machine. Then, the average relative importance ($\mathrm{Imp}_{i,j}$, expressed in percentage) of an input parameter ($i$) in the prediction of a given output feature ($j$) is computed as the mean of the $k=100$ individual absolute SHAP values, such as:

 \begin{equation}
 \mathrm{Imp}_{i,j}=100\frac{\overline{|\mathrm{SHAP}_{i,j}|}}{\sum_k \overline{|\mathrm{SHAP}_{k,j}|}}.
 \end{equation}
Note that we can compute the relative contribution of each input feature in that way because we are working with the standardised input values (mean of zero and standard deviation of one), as otherwise, the SHAP values of the effective temperature will be largely dominant, given its larger spread in numerical values compare to the other input parameters. This also means that these values have to be interpreted with caution, as a change in the scaler method (e.g. from a standard scaler to a min-max scaler) can change the mean absolute SHAP values and so the relative importance of each input feature.

\begin{figure*}
\centering
  \includegraphics[angle=0,viewport=0 45 575 215,clip,width=12cm]{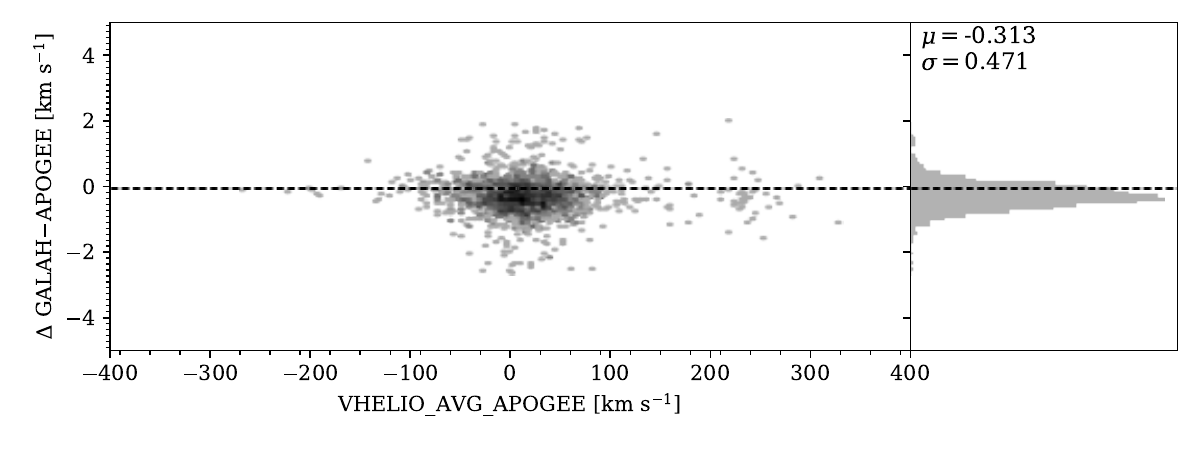}
  \includegraphics[angle=0,viewport=0 0 575 210,clip,width=12cm]{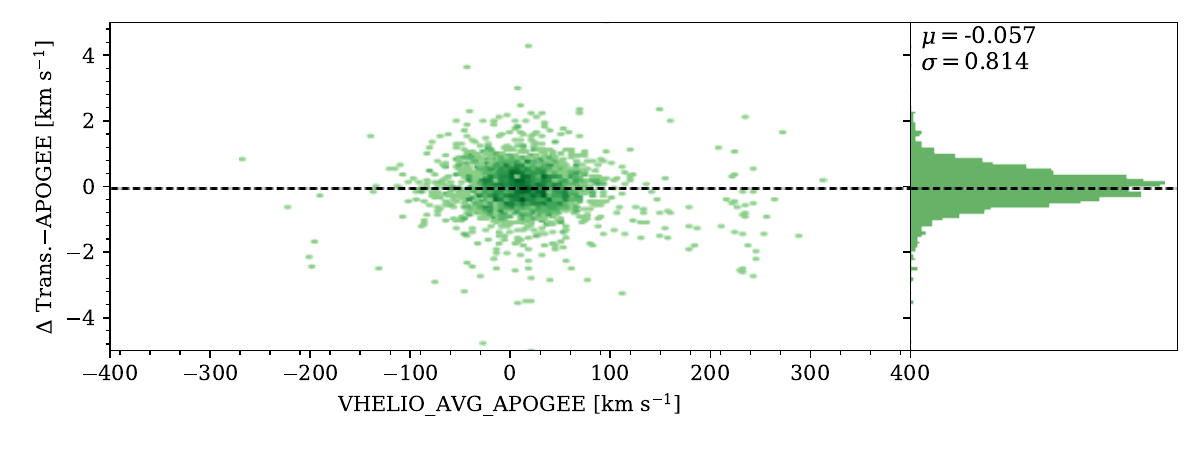}
   \caption{Comparison of the difference of the l.o.s velocity measured by GALAH and APOGEE-~2 on the top panel, and the difference of the velocity transformed by the {\sc SpectroTranslator} algorithm and the APOGEE-~2 data on the bottom panel, as a function of the l.o.s velocity measured by APOGEE-~2. In both cases, the symbols show the stars of the training set. The histograms on the right panels show the distribution of the difference between the l.o.s velocity measured by GALAH and APOGEE-~2 (on top) and between the transformed GALAH l.o.s and the ``original'' APOGEE-~2 values.}
\label{fig:vel}
\end{figure*}
 
The relative importance that each feature has on the transformation from the GALAH to the APOGEE-~2 base for the different spectroscopic parameters are shown in Fig.~\ref{fig:shap_intrinsic}. Before analysing this figure, a few warnings have to be addressed for the reader to not over-interpret it, as SHAP based graphics can lead to misleading interpretation and are not always reliable, as shown by \citet{slack_2020}. First, here we are showing the average relative importance of each feature for the global dataset. However, the importance of each input may depend on the location in the parameter space. For example, the effective temperature plays a less important role in the transformation of the metallicity for the metal-poor stars than for the metal rich stars (see Appendix~\ref{annex:featImp}). This leads to the second point, that the mean of the average SHAP values are computed using a sample of 100 randomly selected stars from the validation sample. This implies that the relative importance of each input feature is mostly indicative of the sub-sample of stars the most present in the samples, i.e. metal-rich stars ([Fe/H]>-0.5) around the main-sequence turn-off. Finally, the relative importance is given for a given number of input parameters. In other words, if we retrain the {\sc SpectroTranslator} algorithm with less input parameters, the precision of the transformation will not necessarily be strongly impacted. For example, in Fig.~\ref{fig:shap_intrinsic}, the two Gaia colours provide 28\% of the information to compute the output effective temperature expressed in the APOGEE-~2 base. However, retraining the algorithm without the colour leads to a prediction $\sim 10$\% less precise on the output effective temperature, with in that case a bigger importance of the input effective temperature. However, with the model presented here, the colours have a lot of weight in the transformation of the effective temperature, as it can penalize stars of a given effective temperature (in the GALAH base) that have Gaia colours different than the average colour of the stars at that temperature. Thus, these stars will have a higher transformation than the others, leading to a larger difference between the input and the output effective temperature. In that way, it might be better to interpret the relative importance of the input features presented in Fig.~\ref{fig:shap_intrinsic} as the weight that each parameter can have to {\it penalize} the transformation GALAH to APOGEE-~2. {\it Therefore, given these different caution points, Fig.~\ref{fig:shap_intrinsic} can and should be used only as an indicative graphics of the relative importance of each input feature.}

\begin{figure*}
\centering
  \includegraphics[angle=0,clip,viewport=0 0 1050 150,width=13cm]{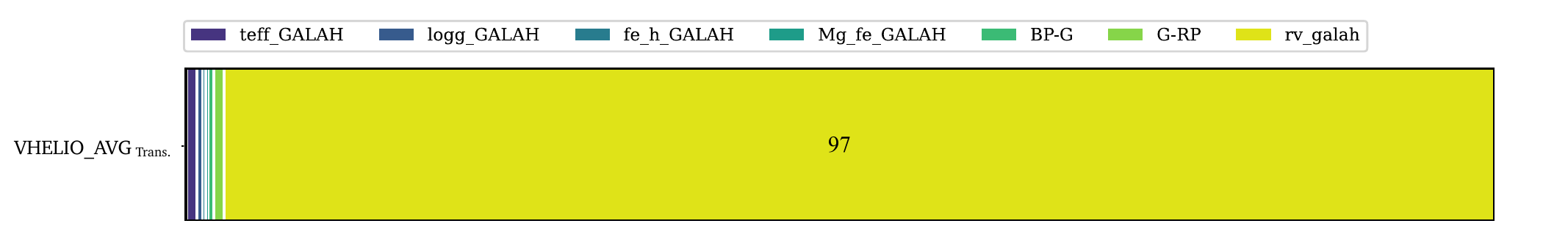}
   \caption{Relative importance of each feature in the transformation of the l.o.s velocity from the GALAH to the APOGEE-~2 base.}
\label{fig:shap_vel}
\end{figure*}

This important caution point made, Fig.~\ref{fig:shap_intrinsic} shows that the transformation of the effective temperature is mostly dependent of the input effective temperature expressed in the GALAH base and of the Gaia colour, but we can see a small dependence on the other parameters, and in particular of the metallicity, as indeed, the difference between the GALAH and APOGEE-~2 effective temperature is more important for metal-poor stars than for the metal rich ones \citep{nandakumar_2022}, due to smaller and less numerous absorption lines present in the former compared to the latter. Similarly, for the surface gravity transformation, the majority of the information comes from the input surface gravity from GALAH (75\%), and the rest is mostly coming from the effective temperature and from the colours, which might be explained by the highest difference between the surface gravity expressed in the GALAH and APOGEE-~2 at the cooler/redder end (T$_\mathrm{eff}<4500$ K) of the training sample than in other region. The same reason is likely behind the relatively high importance of the temperature/colours for the transformation of the metallicity. For the [Mg/Fe] transformation, it is interesting to see that the most important information comes from the metallicity and not from the input GALAH [Mg/Fe]. However, this might be explained by the fact that at the high metallicity end, in the APOGEE-~2 data, there are two [Mg/Fe] tracks clearly identified, usually attributed to the thick and the thin disc, and that at low metallicity ([Fe/H]$<-1.0$), there is only a single track but with a wider distribution of [Mg/Fe], and this separation is less visible with the GALAH data. Therefore, a possible interpretation is that the algorithm first uses the metallicity to have an estimation of [Mg/Fe], and then uses the input [Mg/Fe] from GALAH  to refine its estimation and to break the degeneracy between the thin and thin disc track if the star is metal-rich. Moreover, the dependence on the surface gravity and the effective temperature is likely linked to the fact that 97\% of the metal-poor stars in the training sample are giant stars ($\log$(g)$<3.5$). 

\begin{table*}[h]
 \centering
  \caption{Mean [Fe/H] and [Mg/Fe] derived for the 4 globular clusters studied here using the APOGEE-~2, the ``original'' GALAH, and the translated GALAH into APOGEE-~2 data. The numbers in parentheses correspond the standard deviation of the distribution. Note that the number of stars is the same in GALAH and in the translated datasets. }
   \label{tab:GC}
  \begin{tabular}{@{}lccc@{}}
  \hline
   Cluster & Catalogue(No) &[Fe/H] & [Mg/Fe] \\
 &  & Mean (std dev) & Mean (std dev) \\
    \hline
    \vspace{0.05cm}\\
        & GALAH (191)& -0.72 (0.09) & 0.35 (0.10) \\
NGC 104 & {\sc SpectroTranslator} & -0.75 (0.08) & 0.33 (0.04) \\
        & APOGEE (223)& -0.74 (0.04) & 0.34 (0.03) \\
    \vspace{0.1cm}\\
        & GALAH (16)& -1.08 (0.05) & 0.22 (0.08) \\
NGC 288 & {\sc SpectroTranslator} & -1.21 (0.04) & 0.30 (0.03) \\
        & APOGEE (31)& -1.27 (0.06) & 0.27 (0.03) \\
    \vspace{0.1cm}\\
        & GALAH (11)& -1.01 (0.08) & 0.12 (0.08) \\
NGC 362 & {\sc SpectroTranslator} & -1.15 (0.04) & 0.21 (0.05) \\
        & APOGEE (42)& -1.12 (0.05) & 0.09 (0.06) \\
    \vspace{0.1cm}\\
        & GALAH (9)& -2.02 (0.05) & 0.15 (0.07) \\
NGC 6397 & {\sc SpectroTranslator} & -1.99 (0.04) & 0.32 (0.02) \\
        & APOGEE (116)& -1.92 (0.09) & 0.27 (0.07) \\
    \vspace{0.05cm}\\

\hline
\end{tabular}
\end{table*}

\subsection{Results of the extrinsic network} \label{sec:ext_net}

As visible on the upper panel of Fig.~\ref{fig:vel}, the difference between the velocity measured by GALAH and by APOGEE-~2 are in average offset by $0.3~{\rm km}\,{\rm s}^{-1}$ with a typical scatter of $0.47~{\rm km}\,{\rm s}^{-1}$. However, this velocity difference presents several inhomogeneities, in particular around $\mathrm{VHELIO\_AVG\_APOGEE}=0~{\rm km}\,{\rm s}^{-1}$, where the difference is higher than in other regions. The offset is smaller than the one of $0.52~{\rm km}\,{\rm s}^{-1}$ measured by \citet{tsantaki_2022}. This difference can be explained by the difference in the data release adopted, as \citet{tsantaki_2022} used APOGEE-~2 DR16 and GALAH DR2. It can also be a consequence of the method used to compute the offset, as in \citet{tsantaki_2022}, the offsets are given w.r.t Gaia-RVS, using all the stars in common between a given survey and Gaia-RVS, while in our case, we are only using the stars in common between the two surveys.

By comparing the lower to the upper panel of Fig.~\ref{fig:vel}, we can see that the {\sc SpectroTranslator} algorithm is able to correct for most of this bias, but at a cost of a scatter 1.7 times larger than the original data. The fact that the {\it extrinsic} network performs less well than the {\it intrinsic} one is likely a consequence of the high concentration of stars between $\mathrm{VHELIO\_AVG\_APOGEE}=-50~{\rm km}\,{\rm s}^{-1}$ and $50~{\rm km}\,{\rm s}^{-1}$ compared to other regions. Therefore, the relation learned by the network is largely influenced by the stars located in this region and less for the stars with different velocities. It might be possible to rebalance the influence of each star by imposing that the relative weight on the loss function of the stars at large velocity is more important than that of the stars between $\mathrm{VHELIO\_AVG\_APOGEE}=-50~{\rm km}\,{\rm s}^{-1}$ and $50~{\rm km}\,{\rm s}^{-1}$, either by increasing their number with a Monte Carlo sampling, or directly by including the weights in the loss function. However, our exploratory tests show that the criteria and the way to perform this rebalancing may strongly influence the results, and it is highly connected to how the boundaries of the parameter space of the training/validation sample are defined. As a consequence, we reserve the exploration of which method is the most suitable for a future work.

It is interesting to see that the distribution of the residuals as function of the velocity is different between the upper and lower panels, indicating that the {\sc SpectroTranslator} not only corrects from the bias in velocity, but also finds some correlation between the difference of velocity of the two surveys and some other parameters. In Fig.~\ref{fig:shap_vel}, we can see that the most important parameter in the transformation is, without surprise, the input l.o.s velocity from GALAH, with minor contribution of the effective temperature, the colours, the surface gravity, the metallicity and the Mg abundance respectively. This is very interesting, since \citet{tsantaki_2022} found that both surveys have a trend in metallicity for the l.o.s velocity compared to Gaia, but only APOGEE-~2 has a trend in temperature. However, this can be explained by the fact that we are comparing GALAH to APOGEE-~2, and not to a homogenised catalogue such as done by the SoS. We reserve this comparison for future work.

\subsection{GALAH transformed to APOGEE-~2 catalogue}

We applied the trained {\sc SpectroTranslator} algorithm to the $\sim 590,000$ stars from GALAH DR3 that have a Gaia DR3 counterpart. Note here that we did not apply any selection criteria contrary to the selection made in Sect.~\ref{sec:GALAH}.

We trained 5 times the {\sc SpectroTranslator} by shuffling the training/validation sets. This set of five trained networks is used to estimate the systematic error on each of the transformed parameters caused by the method itself, and to limit the problem of overfitting. A similar method has been used by \citet{thomas_2019} to estimate the  systematic error in the prediction of photometric distances. In practice, the transformed values are obtained using five different machine learnings. Then, the two extreme predictions for each transformed parameter of a given star are discarded. The value of the transformed parameters are given by the mean of the values for the three non-discarded networks, while the standard deviation is considered as the systematic error. 

Another source of uncertainty on the transformed parameters is caused by the measurement uncertainties on each of the spectroscopic parameters. The probabilistic distribution function (PDF) of the transformed parameters is obtained by applying the method described above to a set of 100 Monte-Carlo resampling of the input parameters ($\bf{X_A}$, \boldmath$\theta$\unboldmath). In the catalogue available online\footnote{\url{https://research.iac.es/proyecto/spectrotranslator/}.} we provide the 5, 16, 50, 84, 95-th percentiles of the PDF for each parameter. The systematic error included in the catalogue corresponds to the 50-th percentile of the PDF.

\begin{figure*}
\centering
  \includegraphics[angle=0,clip,width=17cm]{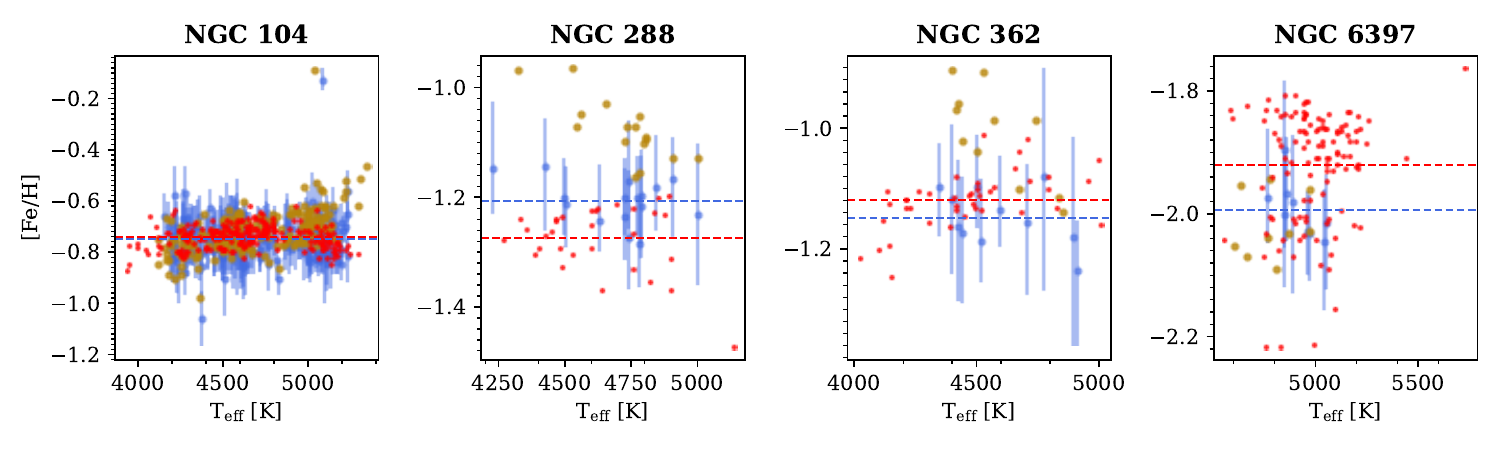}
  \includegraphics[angle=0,clip,width=17cm]{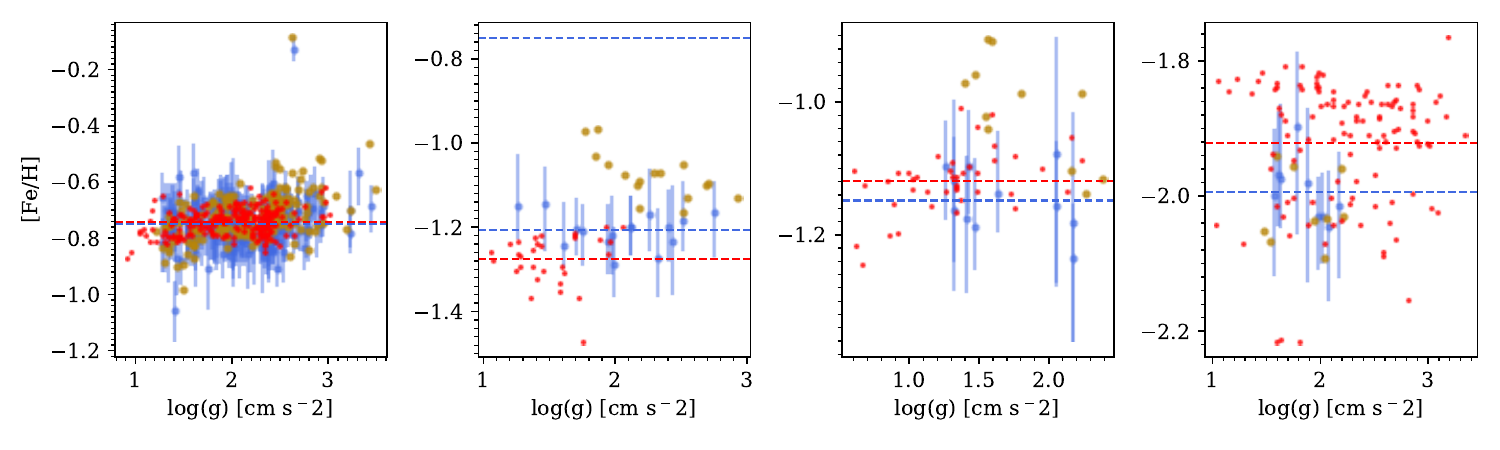}

   \caption{[Fe/H] as function of the effective temperature (top row) and surface gravity (lower row) for 4 globular clusters. The parameters from the ``original'' GALAH data are shown by the orange points, while the value transformed on the APOGEE-~2 base by the {\sc SpectroTranslator} are shown by the blue circles. The red points show the values for the stars present in the APOGEE-~2 DR17 dataset. We emphasize these are not necessarily the same stars as those observed by GALAH. The horizontal red and blue lines indicates the mean metallicity of the cluster measured using the ``original'' APOGEE-~2 and transformed GALAH values, respectively.}
\label{fig:GC_FEH}
\end{figure*}

\begin{figure*}[!ht]
\centering
  \includegraphics[angle=0,clip,width=17cm]{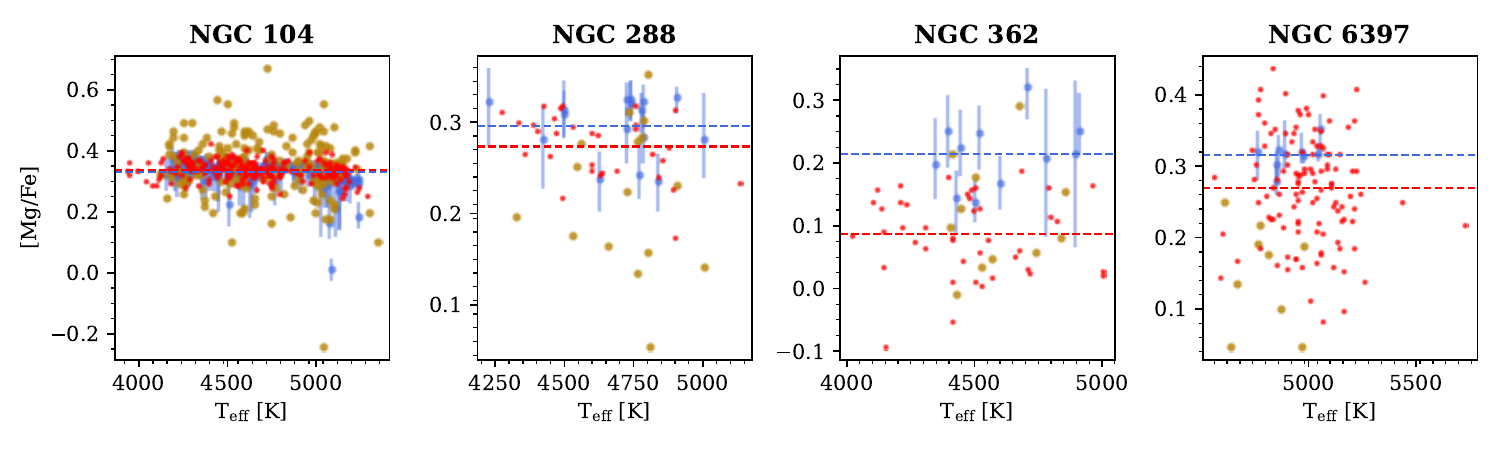}
  \includegraphics[angle=0,clip,width=17cm]{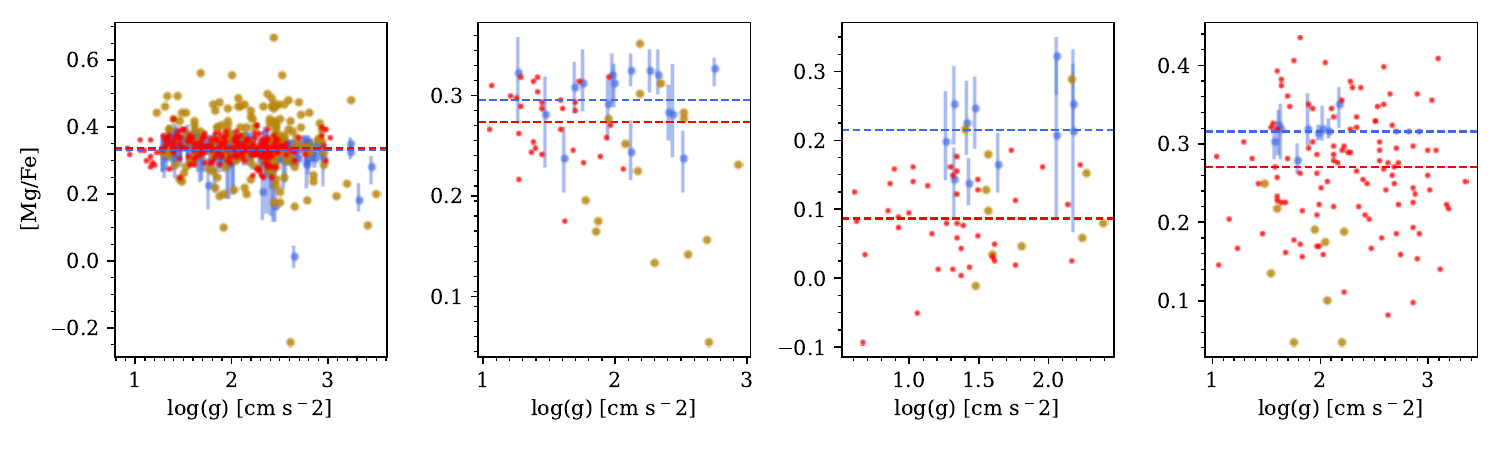}
   \caption{Same as Fig.~\ref{fig:GC_FEH} but for [Mg/Fe] instead of the metallicity. }
\label{fig:GC_MGFE}
\end{figure*}

\begin{figure}[]
\centering
  \includegraphics[angle=0,clip,width=7cm]{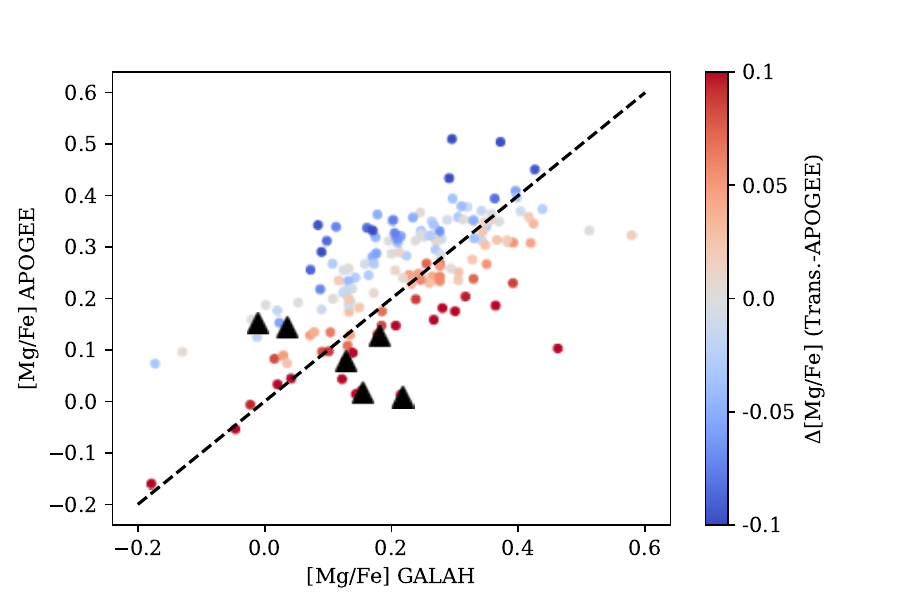}
   \caption{[Mg/Fe] values measured by APOGEE-~2 and GALAH for the six stars of NGC~362 observed by both surveys (in black triangle). The points show the distribution of stars from the training/validation sample in the same range of temperature, surface gravity and metallicity than the stars of NGC~362. They are colour coded by the difference between the [Mg/Fe] transformed by the {\sc SpectroTranslator} and the ``original'' APOGEE-~2.  The dashed line shows the 1:1 relation in the [Mg/Fe] measurement between APOGEE-~2 and GALAH.}
\label{fig:NGC362}
\end{figure}

14\% of the stars from the GALAH DR3 catalogue lack measurements for at least one of the input parameters, particularly [Mg/Fe]. In such cases, we set the missing \textit{renormalized} input value to 0 and proceed with the transformation using this value. Since all input parameters of a network are normalized to have a distribution with a mean of zero and a standard deviation of one with respect to the training sample (see Sect.~\ref{sec:setup}), setting a missing value to 0 is equivalent to assigning it the average value of the parameters in the physical (non-renormalized) space. In such instances, a flag indicating that an input was missing is raised. The provided catalogue includes a flag for both the missing input of the intrinsic and extrinsic networks.

Furthermore, for both networks, the catalogue provides quality flags that indicate if the input/output parameters are inside the range of application of the {\sc SpectroTranslator} as we define it in Sect.~\ref{sec:quality}.

The metadata of the GALAH DR3 catalogue transformed onto the APOGEE-~2 DR17 base are explained in Table.~\ref{tab:descrip_cat}.

\subsection{Validation with globular clusters} \label{sec:GCs}

\begin{figure*}[!ht]
\centering
  \includegraphics[angle=0,clip,width=18.0cm]{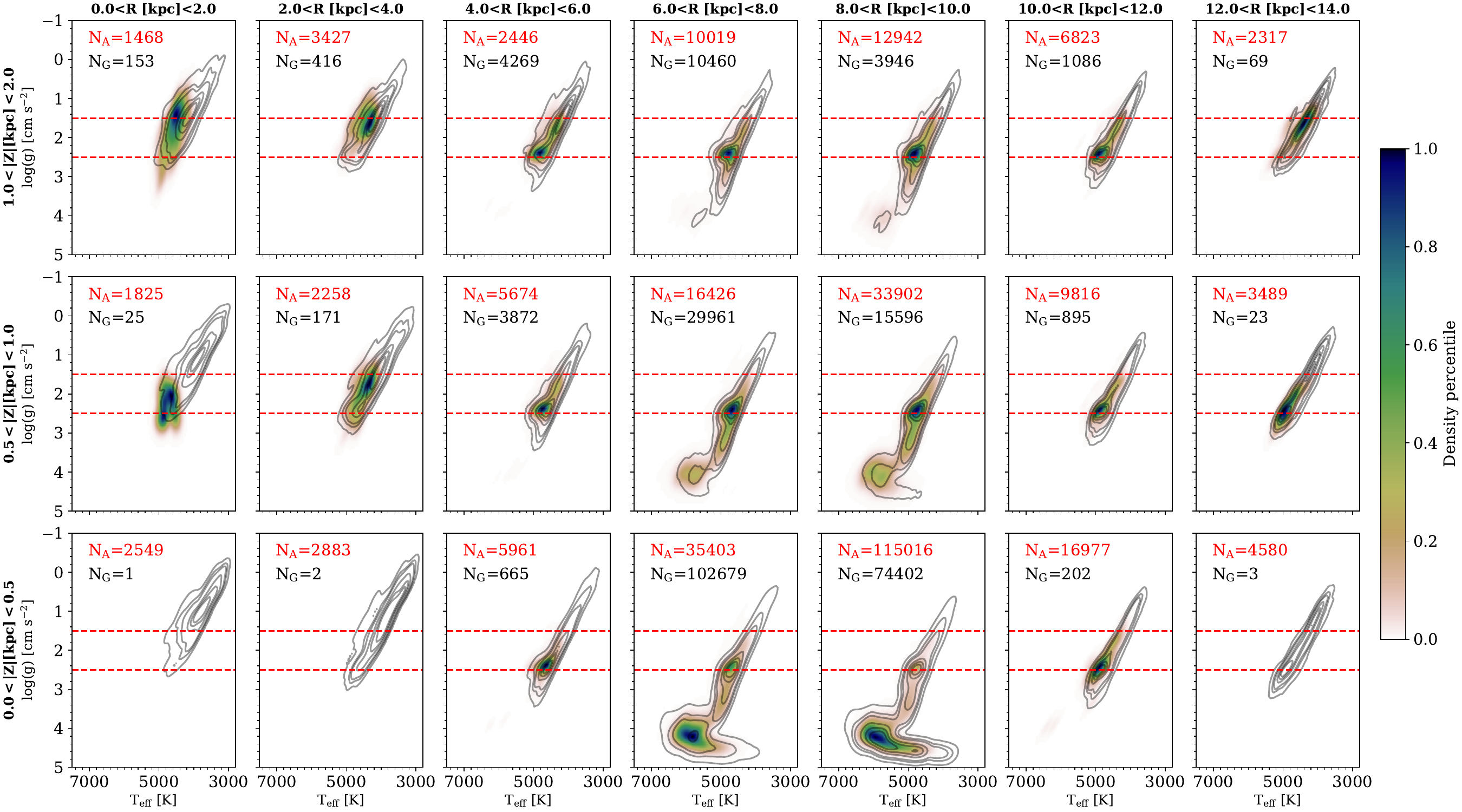}
   \caption{Kiel diagram for different ranges of Galactocentric radii and vertical elevations from the midplane. The 2d histogram shows the relative distribution (made with a kernel density estimator) of the transformed GALAH data in each spatial bin. The grey iso-density contours are plotted
at the 1, 5, 10, 30, 50 and 70\% of maximum density for the stars from APOGEE-~2. In each bin, N$_\mathrm{G}$ and N$_\mathrm{A}$ refer to the number of stars from the GALAH and APOGEE catalogues, respectively. The horizontal dashed red lines show the upper and lower limit for the selection of giant stars used in Sect~\ref{sec:sciencecase}.}
\label{fig:TEFF_LOGG_RZ}
\end{figure*}

\begin{figure*}[!ht]
\centering
  \includegraphics[angle=0,clip,width=18.0cm]{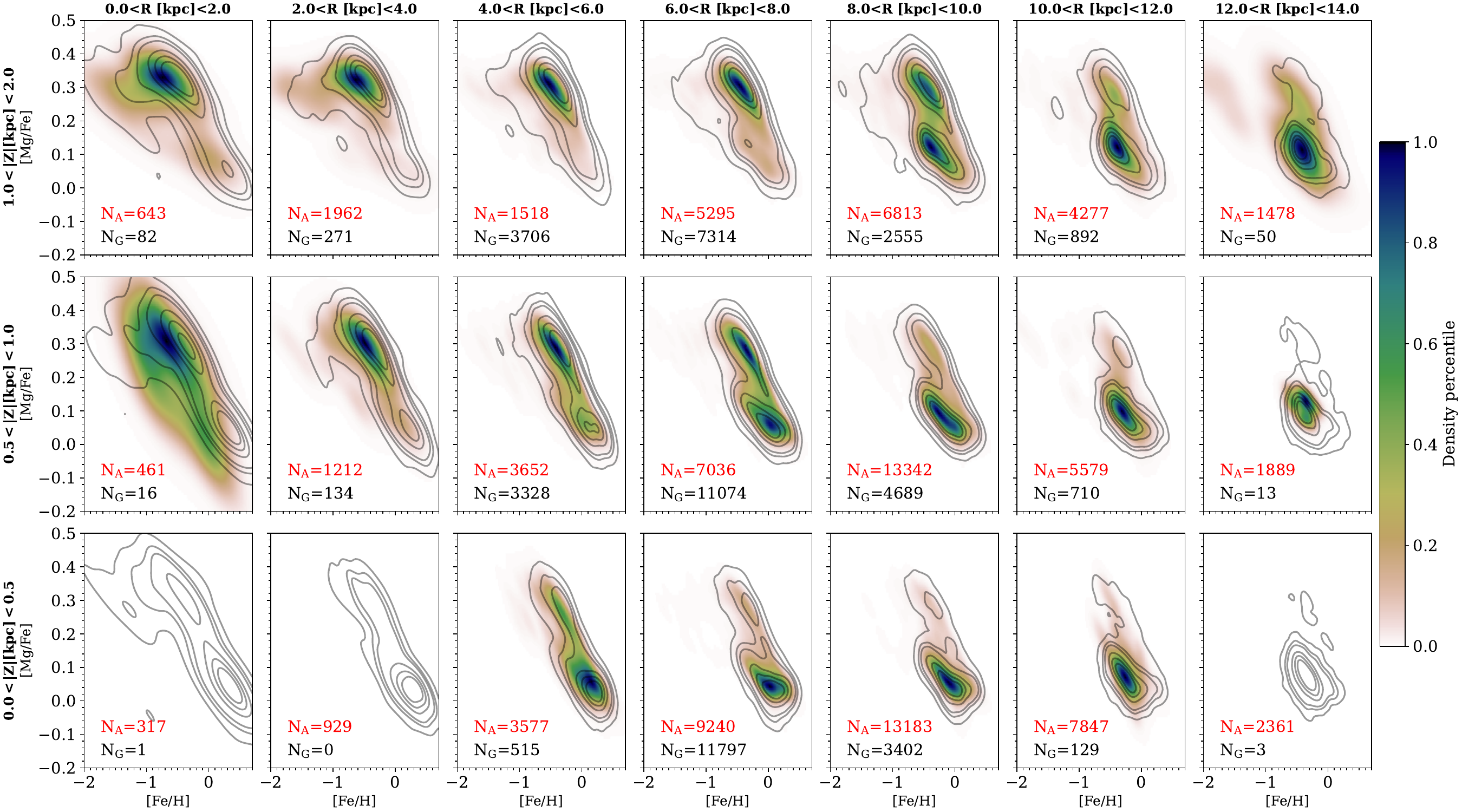}
   \caption{[Mg/Fe] versus [Fe/H] distribution of the selected giant stars in the same spatial bins as for  Fig.~\ref{fig:TEFF_LOGG_RZ}.}
\label{fig:FeH_MgFe_RZ_giants}
\end{figure*}
In this section, our focus is on validating the accuracy of transforming stellar parameters from the GALAH to the APOGEE-~2 base. To achieve this, we use four globular clusters within the GALAH and APOGEE-2 footprint—NGC~104 (47~Tucanae), NGC~288, NGC~362, and NGC~6397—each containing more than one star with a membership probability above 0.5, as determined by the criteria outlined in \citet{vasiliev_2021}. 

To select cluster stars, we apply the criteria detailed in Sections~\ref{sec:GALAH} and \ref{sec:APOGEE} on the respective GALAH and APOGEE-~2 catalogues. Additionally, for the GALAH dataset, we require stars to have correct input and output quality flags for the intrinsic network ({\sc Qflag\_Input\_intrinsic=True} and {\sc Qflag\_Output\_intrinsic=True}), and no missing inputs ({\sc Flag\_missing\_inputs\_intrinsic=False}), ensuring the use of stars with accurate translations (see Table~\ref{tab:descrip_cat}).

Table~\ref{tab:GC} presents the average and standard deviation of metallicity and Mg-abundance obtained using GALAH, APOGEE-~2, and the translated GALAH-into-APOGEE values for each cluster. As expected, the translated GALAH values align more closely with the APOGEE-~2 measurements than the ``original'' GALAH values for both metallicity and Mg-abundance. Notably, the {\sc SpectroTranslator} reduces the scatter found in the ``original'' GALAH data for [Fe/H] and [Mg/Fe] to a value similar to the scatter measured in the APOGEE-~2 sample.

Fig.~\ref{fig:GC_FEH} and \ref{fig:GC_MGFE} illustrate the relationship between [Fe/H] and [Mg/Fe] with $T_\mathrm{eff}$ and log(g). The disparity in the average metallicity measured with APOGEE-~2 and the translated GALAH data, indicated by the horizontal lines, is attributed to the generally broader coverage in effective temperature and surface gravity of the APOGEE-~2 sample. However, in the regions where both surveys overlap, the translated [Fe/H] values are closer to the APOGEE-2 values than the ``original'' GALAH values, particularly for NGC~288 and NGC~6397.

For NGC~362, the translated [Mg/Fe] values exceed those measured by APOGEE-~2 in the same temperature and surface gravity range as the GALAH sample, yet they remain consistent within 1$\sigma$. This is intriguing, considering that the ``original'' GALAH measurements, on average, align more closely with APOGEE-~2 values but exhibit a wider scatter. Six stars are in common between the APOGEE-~2 and GALAH samples, and are therefore part of the training/validation samples). We show on Fig.~\ref{fig:NGC362} [Mg/Fe] values measured by GALAH and APOGEE-~2 for these six stars and we compared their location with stars from the training in the same range of temperature ($4300<\mathrm{T_{eff}}<5000$~K), surface gravity ($1.2<\mathrm{\log{g}}<2.3$~dex) and metallicity ($-1.3<\mathrm{[Fe/H]}<-1.8$~dex). It is clear that four out of the six stars are located in the region where the {\sc SpectroTranslator} overestimates the values of [Mg/Fe] compared to APOGEE-~2. This discrepancy arises because these stars deviate from the GALAH-APOGEE-~2 [Mg/Fe] trends observed in other stars within similar temperature, surface gravity, and metallicity ranges. Specifically, the APOGEE-~2 [Mg/Fe] for these stars is lower (by approximately $-0.13$ dex) than the general trend derived from stars with the same GALAH [Mg/Fe] measurement. It is however, not clear why the [Mg/Fe] values measured by GALAH and APOGEE-~2 in this cluster is different than for the other stars located in the same parameter space region. In particular, it is interesting to note that for NGC~288, which has similar properties to NGC~362, the translated values of [Mg/Fe] are closer to the values from APOGEE-2,showing that the stars of this cluster are similar to the general trend. Nevertheless, globular clusters are very complex environments, with many of them having multiple populations \citep[e.g.][]{bastian_2018,gratton_2019,meszaros_2020}. For instance, it has been observed that a correlation exist between Mg and Al in many clusters \citep[][and references within]{bastian_2018,gratton_2019}, although this correlation is not systematic, especially in clusters rich in metals \citep{pancino_2017}, as are NGC~288 and NGC~362. 

We did not include NGC~5139 ($\omega$-Cen) in this analysis because it has a metallicity scatter more than two time higher than other clusters \citep{meszaros_2020,meszaros_2021}, and has such is less informative of the performance of the {\sc SpectroTranslator} than the other clusters. Nevertheless, it is worth mentioning that for the stars of this cluster observed by GALAH and APOGEE-~2, we found that the {\sc SpectroTranslator} change the average [Fe/H] measured with the GALAH values from $-1.54$~dex to $-1.62$~dex and the average [Mg/Fe] from $0.16$~dex to $0.26$~dex, closer to the average measurement using the APOGEE-~2 values of [Fe/H]$=-1.62$~dex and [Mg/Fe]$=0.27$~dex, respectively.

It will be interesting to study the performance of the \textsc{SpectroTranslator} when the stars belonging to globular clusters are excluded from the training sample. In the current sample, 205 stars belong to globular clusters, the majority (62\%) from NGC~5139 ($\omega$-Cen) and from (23\%) NGC~104 (47~Tucanae). We will explore this, along with the effect of applying a weighting scheme to the training sample, in a dedicated paper in the future.

In summary, with the exception of the [Mg/Fe] measurement in NGC~362, the [Fe/H] and [Mg/Fe] obtained by the {\sc SpectroTranslator} are closer to the APOGEE-~2 measurements and exhibit lower scatter compared to the ``original'' GALAH values. This aligns with expectations for globular clusters \citep[e.g.,][]{masseron_2019,meszaros_2020} which shows small scatter in metallicity.

\section{The 2D-distribution of [Fe/H] and [Mg/Fe] in the Milky Way} \label{sec:sciencecase}

In this section, we showcase the scientific utility of homogenization on a common base facilitated by the {\sc SpectroTranslator}. Our focus is on exploring the insights gained by merging transformed GALAH data with APOGEE-~2 data, particularly to address data gaps in regions not covered by the latter.

The combined catalogue consists of stars from both surveys, selected based on the criteria outlined in Sections~\ref{sec:GALAH} and \ref{sec:APOGEE}. As in the previous section, we ensure the use of stars with accurately translated parameters from the GALAH sample by retaining only those with good input and output quality flags for the intrinsic network, and with no missing inputs ({\sc Qflag\_Input\_intrinsic=True}, {\sc Qflag\_Output\_intrinsic=True}, and {\sc Flag\_missing\_inputs\_intrinsic=False}). For stars observed by both surveys, we preserve the original spectroscopic values from the APOGEE-~2 data.

Note that, to ensure reliability, we discard all stars with non-null {\sc StarHorse\_OUTPUTFLAGS} and remove those within 5 half-light radii and within $\pm 0.5$ kpc from any globular clusters, following the parameters listed in \citep{harris_1996,harris_2010}. Finally, the stars listed as member of a globular cluster or stellar stream in the APOGEE-~2 catalogue and in the catalogue of \citet{schiavon_2024} are also removed. The resulting merged catalogue comprises 571,696 stars, with 56\% from APOGEE-~2 and 44\% from GALAH.

The Cartesian galactocentric coordinates are computed with the {\sc Astropy SkyCoord} package \citep{theastropycollaboration_2018} using the {\sc Starhorse} heliocentric distances from \citet{queiroz_2023}. In this galactocentric frame the Sun is located at [X$_\odot$, Y$_\odot$, Z$_\odot$] = [-8.122~kpc, 0.0~kpc, 20.8~pc].

\begin{figure*}
\centering
  \includegraphics[angle=0,clip,width=18.0cm]{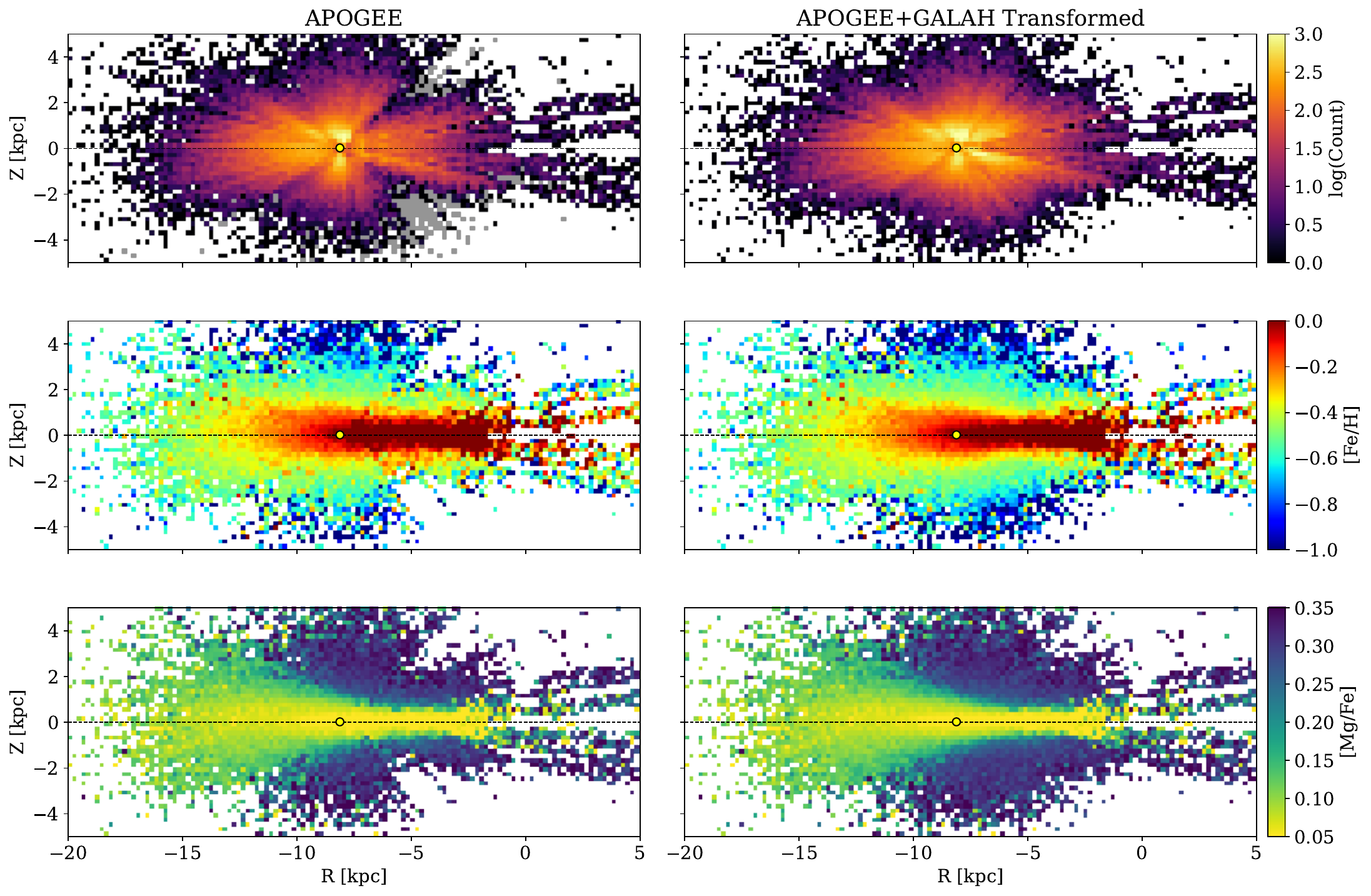}
   \caption{Edge-on view of the global maps of the Milky Way, showing the number of stars observed (upper panels), the median [Fe/H] (middle panels) and median [Mg/Fe] (lower panels) distribution for the APOGEE-~2 sample only (left side), and combined to the transformed GALAH data (right side) in $0.2~\mathrm{kpc} \times 0.2~\mathrm{kpc}$ bins. On the upper left panel, the grey area illustrates the region not covered at all by the APOGEE-~2 survey, but that has been observed by the GALAH survey. In each panel, the dashed black line shows the Galactic mid-plane, and the yellow circle indicates the location of the Sun. Note that R preserve the sign of the X-axis to show the opposite side of the Galaxy.}
\label{fig:Map_APOGEE_GALAH}
\end{figure*}

\begin{figure*}
\centering
  \includegraphics[angle=0,clip,width=13.0cm]{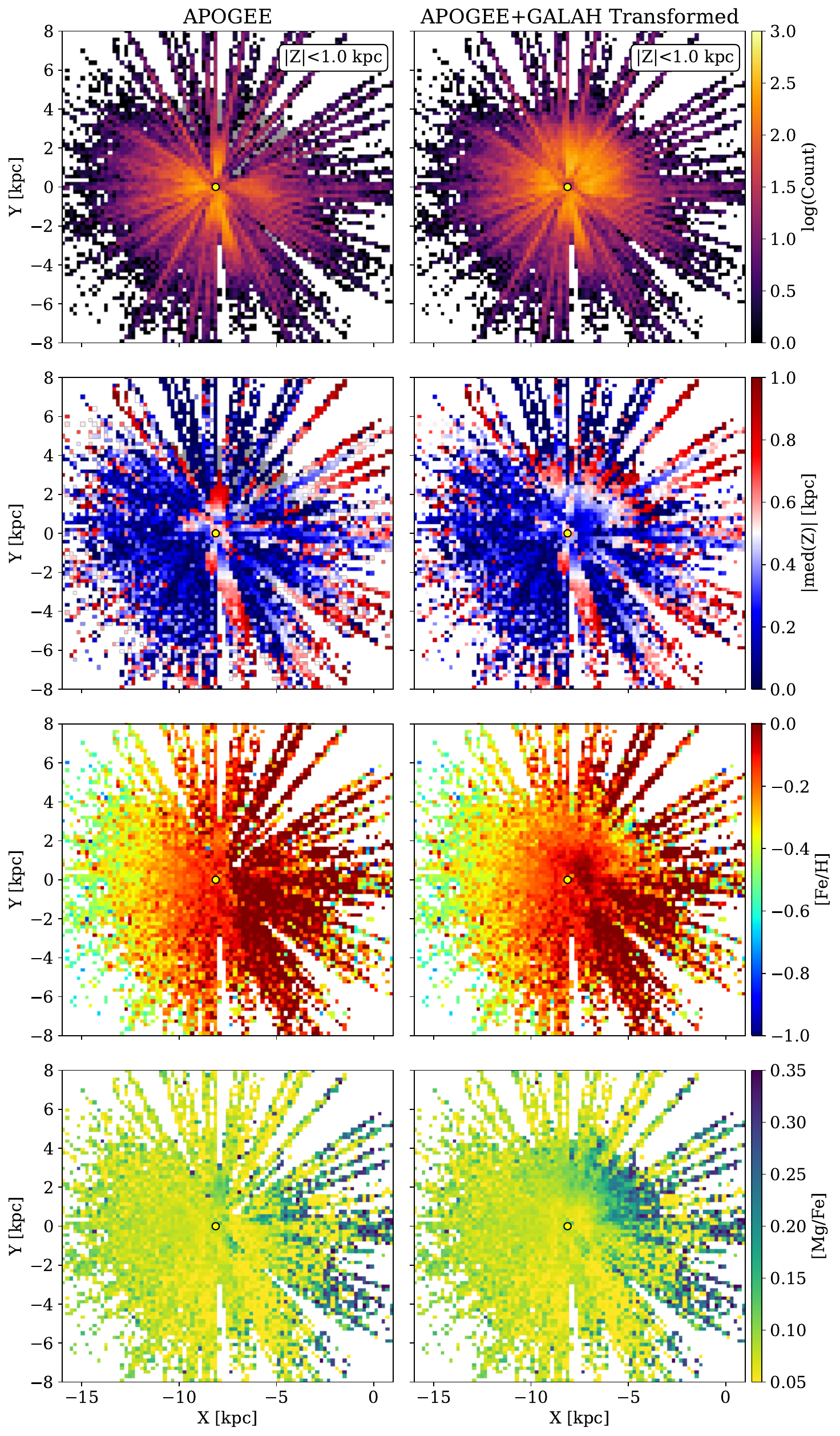}
   \caption{Face-on view of the global maps of the Milky Way of the stars ($|$Z$|<1.0$~kpc), showing the number of stars observed (upper row), the absolute value of the median elevation from the plane of the disc (second row), the median [Fe/H] (third row) and median [Mg/Fe] (lower row) distribution for the APOGEE-~2 sample only (left side), and combined to the transformed GALAH data (right side) in $0.2~\mathrm{kpc} \times 0.2~\mathrm{kpc}$ bins. On the two uppermost left panels, the grey area illustrates the region not covered at all by the APOGEE-~2 survey, but that has been observed by the GALAH survey. In each panel, the yellow circle indicates the location of the Sun.}
\label{fig:Map_azimuth}
\end{figure*}
\subsection{Distribution of [Fe/H] versus [Mg/Fe] across the Galaxy}

We aim in this section to present the distribution of [Mg/Fe] versus [Fe/H] at different Galactocentric radii (R) and vertical elevations from the midplane (Z) in a similar way than \citet{hayden_2015} and \citet{queiroz_2020}. However, the mix of stellar type observed by APOGEE-2 and GALAH change drastically across the Galaxy, but also between the two surveys for a given R and Z. This is visible in Fig.~\ref{fig:TEFF_LOGG_RZ}\footnote{The figure was made using the kernel density estimator included in the python package {\sc sckit-learn} \citep{pedregosa_2011}}, where is presented the Kiel diagrams of the stars from the transformed GALAH catalogue, and from the APOGEE-~2 catalogue at different Galactocentric cylindrical radii (R) and vertical elevations from the midplane (Z).

Therefore, to avoid having artificial variations in the [Mg/Fe] versus [Fe/H] distribution reflecting the underlying variation in the mix of stellar types observed by the two surveys, we decided to restrict our study using only giant stars in the range of surface gravity $2.5>\log(g)>1.5$ dex, as indicated by the two red lines in Fig.~\ref{fig:TEFF_LOGG_RZ}, as they are present at all distances and in both APOGEE-~2 and the transformed GALAH samples. This leads to a selection of 155,885 giant stars from the combined APOGEE-~2 and transformed GALAH sample (66\% from APOGEE and 34\% from GALAH). 

A close match in the distributions on the [Mg/Fe] vs [Fe/H] plane is evident, underscoring the effective performance of the {\sc SpectroTranslator}. However, upon closer examination, subtle differences emerge.

For instance, in the $0<\mathrm{R [kpc]} <2$ and $0.5<\mathrm{|Z| [kpc]}<1.0$ bin, the chemically defined thin disc (low-Mg blob) visible in the APOGEE-~2 data is significantly less visible in the GALAH data. This discrepancy between the two datasets can be attributed to the statistical fluctuations due to the low number of star from the GALAH dataset in that region. Nonetheless, it is interesting to see that the bimodal [$\alpha$/Fe] distribution (which include Mg) observed by APOGEE in the centre of the Milky Way \citep{rojas-arriagada_2019,queiroz_2020} is also visible with the translated GALAH data. However, contrary to these works, a first visual inspection of that region seems to indicate that there is only a single trend which relates the low-Mg to the high-Mg overdensities, i.e. that there is not a degeneracy of [Mg/Fe] for a given [Fe/H]. This is in line with the observation of \citet{hayden_2015,kordopatis_2015,bensby_2017,zasowski_2019,lian_2020,lian_2021,katz_2021,imig_2023}. These differences observed between various studies using the same data are explained by \citet{katz_2021}, who show that the double sequence is only visible for a couple of elements (including the global [$\alpha$/M] used by \citealt{queiroz_2020}), while for the others (including [Mg/Fe]) they present a single trend (see their Appendix~F.). Note that the gap visible in the distribution of the transformed GALAH parameters around [Fe/H]$\sim -1.0$ in the $12<\mathrm{R}<14$~kpc $1.0<\mathrm{|Z|}<2.0$~kpc bin is the combined consequence of the low number of GALAH stars in that bin, and of kernel density estimator method used to make these plots. 

Another notable difference is that the  high-[Mg/Fe] plateau reaches lower [Mg/Fe] values for the translated GALAH sample compared to the APOGEE-~2 sample. This discrepancy stems from the lower accuracy of the {\sc SpectroTranslator} at high-[Mg/Fe] values, as explained in Section~\ref{sec:res_intrinsic}. Furthermore, one can also observe that knee in the [Mg/Fe] vs [Fe/H] distribution generally appears at lower [Fe/H] in the GALAH sample than in APOGEE-~2, although this is not always the case (i.e., in the $6<\mathrm{R}<8$~kpc, $0.5<\mathrm{|Z|}<1.0$~kpc bin). In the bins affected by this discrepancy, we can observe that the distribution of the two surveys on the Kiel diagram is quite different, even for the giant sample used here, with the APOGEE-~2 sample reaching lower temperatures than GALAH for a given $\log(g)$. On the contrary, in the bins where the discrepancy is not visible, we can see that the distributions on the Kiel diagram are similar in the surface gravity range we selected. This might suggest that the difference of location of the knee between APOGEE-~2 and the transformed GALAH data is the consequence of the intrinsic selection function of the two surveys. Note here that these discrepancies are anyway significantly smaller than those that appear when using the ``original'' GALAH values.

\subsection{[Fe/H] and [Mg/Fe] cartography of the Milky Way disc}

Fig.~\ref{fig:Map_APOGEE_GALAH} and Fig.~\ref{fig:Map_azimuth} show, respectively, the edge-on and face-on distribution of median [Fe/H] and [Mg/Fe] across the disc for the 155,885 selected giants using the APOGEE-~2 sample only (left side), and combined with the transformed GALAH sample (right side). Note here that we impose a maximum elevation from the Galactic midplane of $|\mathrm{Z}|<1.0$~kpc for Fig.~\ref{fig:Map_azimuth}. The contribution of the GALAH data-set in complementing the APOGEE-~2 one is clearly visible on the upper panels of both figures. We can see that, not only the GALAH data increase significantly the number of stars observed near the midplane of the disc and at different azimuth, but also allow to observe areas that are not at all observed by APOGEE-~2, in particular toward the Galactic centre in the Southern Galactic hemisphere (Z$<0$~kpc) and Y$>0~$kpc. This allows to expand the [Fe/H] and [Mg/Fe] maps, and to reduce the fluctuations caused by low number statistics, particularly visible when comparing the APOGEE-~2 only and APOGEE-~2+GALAH [Fe/H] maps near the Galactic plane on Fig.~\ref{fig:Map_APOGEE_GALAH}. It is interesting to notice that the global distribution of [Fe/H] and [Mg/Fe] in Fig.~\ref{fig:Map_APOGEE_GALAH}is not drastically different when using only the APOGEE-~2 data or combined with the transformed GALAH data, and that the area without APOGEE data (highlighted in grey in the upper left panel) does not mark a discontinuity with the rest of the map. This highlights the efficiency of the {\sc SpectroTranslator} in transforming the spectroscopic values from one base to another.

On the contrary, in Fig.~\ref{fig:Map_azimuth}, we can see that the median [Fe/H] and [Mg/Fe] differ when the transformed GALAH values are added to the APOGEE-2 sample. However, these differences reflect the variation in elevation sampled by these stars, as shown in the second row of Figure~\ref{fig:Map_azimuth}. Here, we observe that the median elevation of the region highly populated by GALAH stars is higher than that of the APOGEE stars at the same position in the X-Y plane. This explains why, in these regions, the median [Fe/H] is lower and the median [Mg/Fe] is higher when the GALAH stars are included compared to the case where only the APOGEE-~2 data are used, due to the vertical gradient present in the disc.

Indeed, a qualitative inspection of the edge-on median metallicity map (middle panels of Fig.~\ref{fig:Map_APOGEE_GALAH}) reveals a clear radial and vertical metallicity gradient in the disc, with higher average metallicities near the Galactic centre and closer to the midplane, which decline towards outer radii and higher elevations. In contrast, the median [Mg/Fe] (lower panels of Fig.~\ref{fig:Map_APOGEE_GALAH}) is lower near the Galactic midplane and rapidly increases towards higher elevations, marking the transition between the thin and thick discs. The flaring of the thin disc (low [Mg/Fe]) beyond a radius of 6 kpc is clearly visible in the [Mg/Fe] map. Overall, these maps closely resemble the cartography recently produced by \citet{gaiacollaboration_2023} and \citet{imig_2023} using Gaia and APOGEE-~2 DR17 data, respectively.

\begin{figure*}[!ht]
\centering
  \includegraphics[angle=0,clip,width=18.0cm]{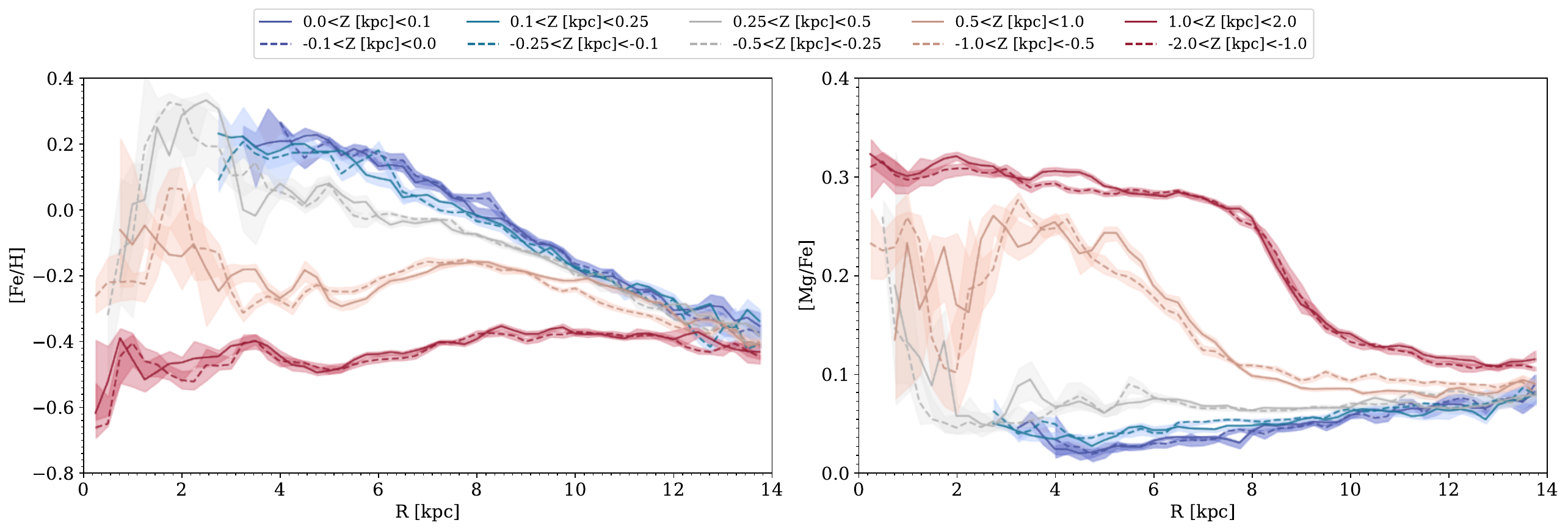}
   \caption{Radial gradients of the metallicity (left) and [Mg/Fe] (right) for
different elevations from the Galactic midplane. The trends are computed as running medians in bins of 0.5 kpc, with a 50 percent overlap, provided that at least 20 stars are available to compute the median. The shaded areas represent the uncertainty on the median (obtained from the 16th and 84th percentile of 1000 bootstrap samples). The continous lines show the trend above the Galactic midplane (Z>0), and the dashed lines the trend below the midplane (Z<0).}
\label{fig:MH_MG_R}
\end{figure*}

\begin{figure*}[!ht]
\centering
  \includegraphics[angle=0,clip,width=18.0cm]{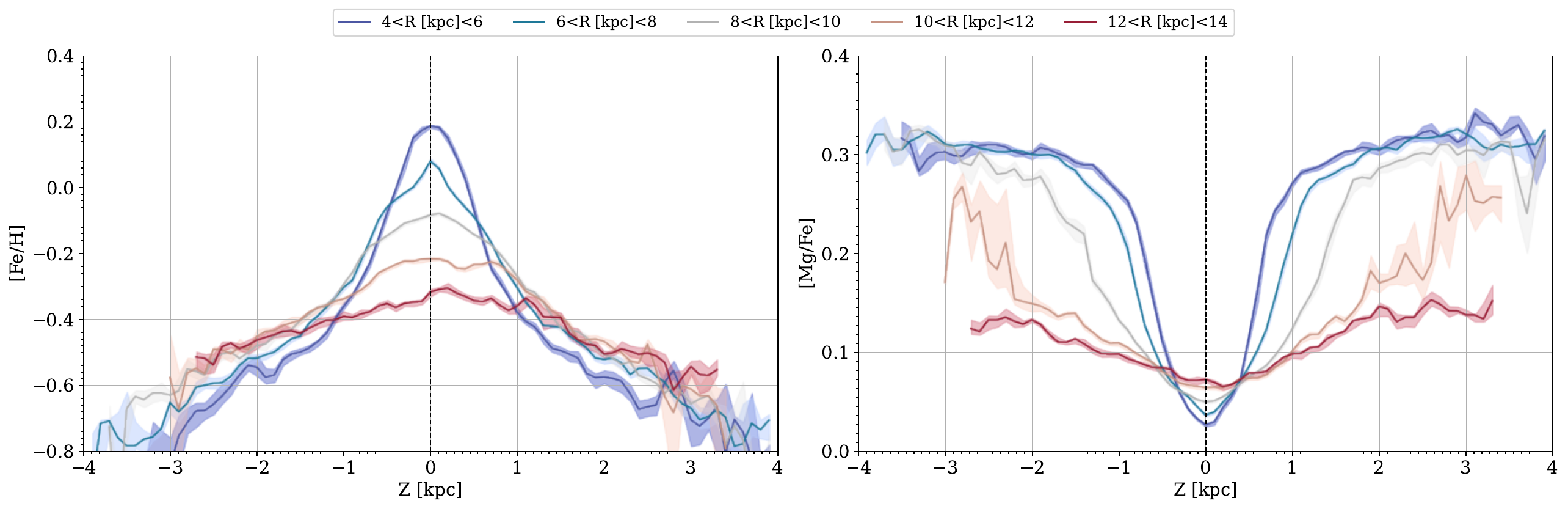}
   \caption{Vertical gradients of the metallicity (left panel) and [Mg/Fe] (right panel) for different radial distance from the Galactic centre. The trends are computed as running medians in bins of 0.2 kpc, with a 50 percent overlap, provided that at least 20 stars are available to compute the median. The shaded areas represent the uncertainty on the median (obtained from the 16th and 84th percentile of 1000 bootstrap samples).}
\label{fig:MH_MG_Z}
\end{figure*}

\subsection{Radial and vertical gradients} \label{sec:gradient}

Here we investigate the overall radial and vertical gradients of [Fe/H] and [Mg/Fe] across the Galaxy. 

We divided the 155,885 selected giants into 5 ranges of vertical elevation on either side of the disc midplane. For each slice, we computed the running median [Fe/H] and [Mg/Fe] as a function of radius in bins 0.5 kpc with a 50\% overlap between each bin. Following the method used by \citet{martig_2016}, the uncertainty on the median value in each bin is estimated by measuring the median for 1,000 bootstrap realizations of the sample. The uncertainties correspond to the 16th and 84th percentiles of this distribution.

Fig.~\ref{fig:MH_MG_R} illustrates the median metallicity (left panel) and [Mg/Fe] (right panel) radial trends for five different elevations above (solid lines) and below (dashed lines) the Galactic midplane. We only show bins containing at least 20 stars. Similar to \citet{gaiacollaboration_2023}, we find that both metallicity and [Mg/Fe] trends exhibit similar behaviors on either side of the Galactic plane, irrespective of the elevation. At most elevations, we observe a break in the metallicity trend around 6-8 kpc at |z|>0.25 kpc, consistent with previous studies using different catalogs \citep[e.g.,][]{haywood_2019, kordopatis_2020, katz_2021, gaiacollaboration_2023}. However, the radius of the break seems to vary with elevation, occurring farther out at higher elevations compared to near the plane of the disc. For regions near the Galactic midplane, no clear break is evident due to limited coverage, though a tentative plateau beginning around $\mathrm{R}=5$ kpc is discernible. This pattern aligns with findings by \citet{imig_2023} when considering both thin and thick disc populations. Notably, the [Fe/H] value of the plateau differs from that found by \citet{gaiacollaboration_2023} ($\mathrm{[Fe/H]}=0.0$ dex versus 0.2 dex here), but the median metallicity converges toward $\mathrm{[Fe/H]}\simeq-0.4$ dex at large radial distances, consistent with various studies \citep[e.g.,][]{eilers_2022,imig_2023} including \citet{gaiacollaboration_2023}. The absence of the $\sim 0.5$ kpc wide wiggles found by \citet{gaiacollaboration_2023} for $|Z| < 0.5$ kpc and $R \sim 8.5$ kpc on both sides of the disc, which they partially attributed to the presence of hot turn-off stars in their sample, suggests that these wiggles may be consequences of their geometric selection biases rather than dynamical effects. 

Regarding the [Mg/Fe] trends, they differ significantly from the [$\alpha$/Fe] trends found by \citet{gaiacollaboration_2023}. For instance, we do not observe a clear break with a transition from a negative to a positive trend around 6 kpc for the region near the Galactic midplane. Instead, we find a clear drop in the median [Mg/Fe] value in the highest elevation bin from $\mathrm{[Mg/Fe]}=0.3$ dex to $\mathrm{[Mg/Fe]}=0.15$ dex at 8 kpc, while \citet{gaiacollaboration_2023} found that [$\alpha$/Fe] declines smoothly. These differences are explained by the dominance of calcium in their [$\alpha$/Fe] abundances, which has a less extended distribution than magnesium, and because calcium is a weaker indicator of the ratio of supernovae type II over type Ia, and thus of the separation between the chemically selected thin and thick discs \citep[e.g.,][]{minchev_2012a,minchev_2012}. Furthermore, we are using data with higher spectroscopic resolution than Gaia-RVS, which contributes to explaining these differences. A less significant drop is also visible at $0.5<|Z|<1.0$ kpc, with a decrease from $\mathrm{[Mg/Fe]}=0.2$ dex to $\mathrm{[Mg/Fe]}=0.1$ dex around $6<\mathrm{R}<8$ kpc. These trends resemble those found by \citet{martig_2016} and are consequences of the flaring of the thin (low-[Mg/Fe]) disc beyond $\sim 6$ kpc. This flaring is likely caused by the radial migration of thin disc stars \citep[e.g.,][]{minchev_2012a, minchev_2015,kordopatis_2015a}. For instance, \citet{minchev_2012a}, show that migrator stars increase the velocity dispersion, and so the scale-height, of the disc outside the corotation radius of the Galactic bar. Assuming a flat rotation speed of $233~\mathrm{km}~\mathrm{s}^{-1}$ \citep{poder_2023} and a bar pattern speed of $\simeq 40~\mathrm{km}~\mathrm{s}^{-1}~\mathrm{kpc}^{-1}$ \citep{wegg_2015, sormani_2015, li_2016,portail_2017,sanders_2019,clarke_2019}, the corotation radius of the Milky Way with the bar is located at 5.8 kpc. Therefore, if the migrated stars come from the thin (low-[Mg/Fe]) disc, we therefore expect to observe an increase of it beyond $\sim$ 5.8 kpc. This aligns with the qualitative observations seen in Fig.~\ref{fig:Map_APOGEE_GALAH} and \ref{fig:MH_MG_R}, where we observe first a slow decrease in median [Mg/Fe] with radius above $|Z|=0.5$ kpc between $\sim$ 6 and 8 kpc, followed by a rapid drop beyond. 

Fig.~\ref{fig:MH_MG_Z} presents the vertical gradient of the median metallicity and [Mg/Fe] for different Galactic radii. Both the [Fe/H] and [Mg/Fe] distributions exhibit stronger vertical gradients near the Galactic midplane compared to \citet{gaiacollaboration_2023}. This difference partly stems from the radii used between both studies, as well as differences in [Fe/H] and [Mg/Fe] values between Gaia and APOGEE-~2. For both [Fe/H] and [Mg/Fe], the vertical gradient is stronger near the Galactic centre than at large radii. We observe a smoother transition between the low and high [Mg/Fe] disc at larger radii, with the median [Mg/Fe] decreasing with radius at a given elevation, for $|Z|>0.5$ kpc, particularly beyond $\sim 8$ kpc. This is a consequence of the flaring of the thin (low-[Mg/Fe]) disc at large distances, as visible in Fig.~\ref{fig:Map_APOGEE_GALAH}. For the metallicity, the median value evolves gradually with radius inside $|Z|<1.0$ kpc. For higher elevations, the evolution of metallicity becomes similar at every radius, except in the inner Galaxy where the median metallicity is lower than at other radii at a given elevation, although it exhibits a similar gradient. Notably, the vertical distribution of the median [Fe/H] is skewed toward positive elevations at large radii, in particular in the outermost radii studied ($12< R < 14$ kpc). This shift is also visible in the metallicity map of Fig.~\ref{fig:Map_APOGEE_GALAH}. The vertical shift toward positive vertical elevation at large distances is also present in the [Mg/Fe] distribution, although it is less significant than in the metallicity distribution. Note here that this asymmetry is not an artifact created by the {\sc SpectroTranslator}, as it is also visible in the APOGEE-~2 data (see Appendix~\ref{sec:grad_APOGEE}).

The observed asymmetry in the median [Fe/H] and [Mg/Fe] distributions is challenging to explain solely by the radial migration of stars, as to our knowledge, the radial migration should increase the scale height symmetrically on both sides of the disc \citep[e.g.,][]{sellwood_2002,minchev_2012a,minchev_2012,minchev_2013,minchev_2014,minchev_2015,minchev_2017,johnson_2021}. Instead, it is likely the consequence of perturbations generated by a satellite galaxy passing through the disc, as proposed by \citep{ibata_1998,velazquez_1999,kazantzidis_2008,villalobos_2008,purcell_2010,gomez_2016}. More recently, \citet{laporte_2018a,laporte_2018b} demonstrated that a coupling between the passage of the Sagittarius dwarf and the dark matter wakes generated by the infall of the Large Magellanic Clouds can induce a vertical asymmetry of density along the midplane of the Galactic disc at large Galactocentric radii. This scenario has been proposed to explain various phenomena such as the high vertical flaring observed in the outer edge of the Milky Way \citep{thomas_2019}, the formation of the Monoceros-Anticentre stream \citep{laporte_2020}, and the presence of phase-space spiral structure in the Solar vicinity \citep{antoja_2018, laporte_2020a}. Furthermore, \citet{ruiz-lara_2020} demonstrated that the three narrow episodes of enhanced star formation in the Milky Way during the last 6~Gyr coincide with the pericentre passages of the Sagittarius dwarf galaxy. Nevertheless, to confirm this scenario, a proper chemo-dynamical analysis is needed, \citep[e.g.][]{binney_2023,binney_2024}. However, such an analysis is beyond the scope of this paper.

\section{Conclusions} \label{sec:conclusions}

In this paper, we presented the \textsc{SpectroTranslator}, a new data-driven algorithm that can transform spectroscopic parameters from the base of one catalogue to that of another catalogue. This algorithm is composed of two deep-residual networks, an \textit{intrinsic} and an  \textit{extrinsic} network. In the former, all input parameters play a role in computing the transformation of the spectroscopic parameters from one base to another. This is mostly used to compute the transformation of fundamental stellar parameters, such as the effective temperature, surface gravity, metallicity, or $\alpha$-abundance. In the second, the transformation of only one of the parameters depends on all the other parameters, but this parameter does not affect the transformation of the others, perfectly adapted to compute the transformation of line-of-sight velocity or of individual abundances. We demonstrated the ability of \textsc{SpectroTranslator} by transforming the effective temperature, surface gravity, metallicity, and [Mg/Fe] of the GALAH DR3 catalogue to the APOGEE-~2 DR17 base using the \textit{intrinsic} network with very high precision, similar to the results obtained by \citet{nandakumar_2022} using the \textit{Cannon} directly on the spectra. Using a method to measure the importance of each parameter in the transformation, we find for example that the metallicity transformation from GALAH to APOGEE-~2 is strongly dependent on the effective temperature and surface gravity, and that the most important parameter for the transformation of [Mg/Fe] between the two surveys is the metallicity. We also transformed the line-of-sight velocity using the \textit{extrinsic} network, although in that case, the algorithm is able to correct most of the biases existing between the two surveys, but it increases the scatter. This shows the limitation in the training of the algorithm that might need to rebalance the weights of each star to find a proper relation between the two surveys. We are going to explore the impact of different rebalance strategy in future work. 

We also presented the scientific potential of this algorithm by measuring the distribution of [Fe/H] and [Mg/Fe] across our Galaxy. Unsurprisingly, we recovered the radial and vertical [Fe/H] and [Mg/Fe] gradient and the low/high-[Mg/Fe] bimodality found in past studies \citep[e.g.][]{hayden_2015,haywood_2018,rojas-arriagada_2019,kordopatis_2020,queiroz_2020,gaiacollaboration_2023,imig_2023,guiglion_2024}. However, contrary to many previous studies using the APOGEE-~2 data, we found that the inner Galaxy present a single trend between the high-[Mg/Fe] and low-[Mg/Fe] region, which is explained by the fact that we are using [Mg/Fe] and not the global [$\alpha$/M] \citep{katz_2021}. Interestingly, we also found that the distribution of [Fe/H] and [Mg/Fe] across the midplane of the MW is asymmetric at large radius, with the northern Galactic hemisphere more metal-rich and [Mg/Fe]-poor compared to the southern hemisphere at the same elevation. We propose that this asymmetry is a consequence of the perturbations of the outer disc generated by the passage of the Sagittarius dwarf galaxy and/or of the dark matter wakes generated by the infall of the Large Magellanic Clouds.

The transformed GALAH to APOGEE-~2 catalogue presented here, the inverse transformation (APOGEE-~2 to GALAH), as well as the training/validation samples used to train \textsc{SpectroTranslator} in both cases, are available on our website \url{https://research.iac.es/proyecto/spectrotranslator/}. We aim to update this website regularly by including the transformation of more spectroscopic catalogues, such as SEGUE \citep{yanny_2009}, LAMOST \citep{zhao_2012a}, Gaia-ESO \citep{randich_2022}, DESI \citep{flaugher_2014,cooper_2023}, Gaia \citep{recio-blanco_2023}, or H3 \citep{conroy_2021}. We will also use this methodology to homogenise spectroscopic parameters including those from the next generation of large spectroscopic surveys, such as WEAVE \citep{dalton_2012,jin_2023} and 4-MOST \citep{dejong_2010} surveys. These two surveys are complementary one to another, as WEAVE will observe the Northern sky, while 4-MOST will observe the Southern sky. Apply the {\sc SpectroTranslator} to these data to homogenize them on the same base will thus allow the access to the entire volume of our Galaxy.

The {\sc SpectroTranslator} algorithm promises to play a crucial role in standardizing various spectroscopic surveys onto a unified basis. This capability is particularly significant given the different spatial coverage of the different large spectroscopic surveys currently underway. By providing a mean to seamlessly translate spectroscopic parameters across different observational bases, the {\sc SpectroTranslator} algorithm facilitates comparative analyses that leverage data from multiple sources. In doing so, it contributes to increasing the legacy of large surveys and in their scientific exploitation, and it is poised to become an indispensable tool to unravel the complexities of our Galactic environment and beyond.

Several updates of the {\sc SpectroTranslator} algorithm are already possible, either by weighting the stars of the training sample, in order to decrease the high heterogeneous distribution of stars across the parameter space, or by transforming the architecture of the network to use a Bayesian Neural Network, that will take better in consideration the uncertainties and possible degeneracy between the input and output features. We strongly encourage researchers to download (see details below) and to experiment with the {\sc SpectroTranslator} algorithm and to collaborate with the authors to improve it.

\section*{Data accessibility}
The {\sc SpectroTranslator} algorithm is publicly available on GitHub \url{https://github.com/GFThomas/SpectroTranslator.git}. The training/validation samples as well as the translated catalogues for different surveys is available on \url{https://research.iac.es/proyecto/spectrotranslator/}.

\section*{Acknowledgements}

G. Thomas, G. Battaglia, E. Fern\'andez-Alvar and C. Gallart acknowledge support from the Agencia Estatal de Investigación del Ministerio de Ciencia en Innovación (AEI-MICIN) and the European Regional Development Fund (ERDF) under grant number PID2020-118778GB-I00/10.13039/501100011033 and the AEI under grant number CEX2019-000920-S.
E.F.A acknowledges the HORIZON TMA MSCA Postdoctoral Fellowships Project TEMPOS, number 101066193, call HORIZON-MSCA-2021-PF-01, by the European Research Executive Agency. 
This project has received funding from the European Research Council (ERC) under the European Union’s Horizon Europe programme "StarDance: the non-canonical evolution of stars in clusters" (ERC-2022-AdG, Grant Agreement 101093572, PI: E. Pancino).
F.G. gratefully acknowledge support from the French National Research Agency (ANR) funded project “MWDisc” (ANR-20-CE31-0004) and “Pristine” (ANR-18-CE31-0017).

This work has made use of data from the European Space Agency (ESA) mission {\it Gaia} (\url{https://www.cosmos.esa.int/gaia}), processed by the {\it Gaia} Data Processing and Analysis Consortium (DPAC, \url{https://www.cosmos.esa.int/web/gaia/dpac/consortium}). Funding for the DPAC has been provided by national institutions, in particular the institutions participating in the {\it Gaia} Multilateral Agreement.

Funding for the Sloan Digital Sky Survey IV has been provided by the Alfred P. Sloan Foundation, the U.S. Department of Energy Office of Science, and the Participating Institutions. SDSS-IV acknowledges support and resources from the Center for High-Performance Computing at the University of Utah. The SDSS web site is \url{www.sdss.org}.
SDSS-IV is managed by the Astrophysical Research Consortium for the Participating Institutions of the SDSS Collaboration including the Brazilian Participation Group, the Carnegie Institution for Science, Carnegie Mellon University, the Chilean Participation Group, the French Participation Group, Harvard-Smithsonian Center for Astrophysics, Instituto de Astrof\'isica de Canarias, The Johns Hopkins University, 
Kavli Institute for the Physics and Mathematics of the Universe (IPMU) / University of Tokyo, Lawrence Berkeley National Laboratory, 
Leibniz-Institut f\"ur Astrophysik Potsdam (AIP),  
Max-Planck-Institut f\"ur Astronomie (MPIA Heidelberg), 
Max-Planck-Institut f\"ur Astrophysik (MPA Garching), 
Max-Planck-Institut f\"ur Extraterrestrische Physik (MPE), 
National Astronomical Observatory of China, New Mexico State University, 
New York University, University of Notre Dame, 
Observat\'ario Nacional / MCTI, The Ohio State University, 
Pennsylvania State University, Shanghai Astronomical Observatory, 
United Kingdom Participation Group,
Universidad Nacional Aut\'onoma de M\'exico, University of Arizona, 
University of Colorado Boulder, University of Oxford, University of Portsmouth, 
University of Utah, University of Virginia, University of Washington, University of Wisconsin, 
Vanderbilt University, and Yale University.

This work made use of the Third Data Release of the GALAH Survey (Buder et al. 2021). The GALAH Survey is based on data acquired through the Australian Astronomical Observatory, under programs: A/2013B/13 (The GALAH pilot survey); A/2014A/25, A/2015A/19, A2017A/18 (The GALAH survey phase 1); A2018A/18 (Open clusters with HERMES); A2019A/1 (Hierarchical star formation in Ori OB1); A2019A/15 (The GALAH survey phase 2); A/2015B/19, A/2016A/22, A/2016B/10, A/2017B/16, A/2018B/15 (The HERMES-TESS program); and A/2015A/3, A/2015B/1, A/2015B/19, A/2016A/22, A/2016B/12, A/2017A/14 (The HERMES K2-follow-up program). We acknowledge the traditional owners of the land on which the AAT stands, the Gamilaraay people, and pay our respects to elders past and present. This paper includes data that has been provided by AAO Data Central (datacentral.org.au).

\bibliographystyle{aa}
\bibliography{./biblio}

\appendix

\section{Metadata of the GALAH data translated onto the APOGEE-~2 base} \label{annex:metadata}

Table~\ref{tab:descrip_cat} shows the description of all the spectroscopic parameters translated from the GALAH onto the APOGEE-~2 base.

\begin{table}[h]
 \centering
  \begin{tabular}{l|l|p{10 cm}}
  \hline
   Column name & Unit / format & Description \\
    \hline
   {\sc sobject\_id} & & GALAH observation identifier \\
   {\sc source\_id} & & Gaia DR3 unique source identifier \\
   & & \\
   {\sc TEFF\_pred\_5} & K & 5-th percentile of the {\sc SpectroTranslator} effective temperature \\
   {\sc TEFF\_pred\_16} & K & 16-th percentile of the {\sc SpectroTranslator} effective temperature \\
   {\sc TEFF\_pred\_50} & K & 50-th percentile of the {\sc SpectroTranslator} effective temperature \\
   {\sc TEFF\_pred\_84} & K & 84-th percentile of the {\sc SpectroTranslator} effective temperature\\
   {\sc TEFF\_pred\_95} & K & 95-th percentile of the {\sc SpectroTranslator} effective temperature\\
   {\sc TEFF\_pred\_ERR} & K & Systematic error of the {\sc SpectroTranslator} on the transformation of the effective temperature\\
   & & \\
   {\sc LOGG\_pred\_5} &  cm~s$^{-2}$ & 5-th percentile of the {\sc SpectroTranslator} surface gravity\\
   {\sc LOGG\_pred\_16} &  cm~s$^{-2}$ & 16-th percentile of the {\sc SpectroTranslator} surface gravity\\
   {\sc LOGG\_pred\_50} &  cm~s$^{-2}$ & 50-th percentile of the {\sc SpectroTranslator} surface gravity\\
   {\sc LOGG\_pred\_84} &  cm~s$^{-2}$ & 84-th percentile of the {\sc SpectroTranslator} surface gravity\\
   {\sc LOGG\_pred\_95} &  cm~s$^{-2}$ & 95-th percentile of the {\sc SpectroTranslator} surface gravity\\
   {\sc LOGG\_pred\_ERR} &  cm~s$^{-2}$ & Systematic error of the {\sc SpectroTranslator} on the surface gravity\\
   & & \\
   {\sc M\_H\_pred\_5} &   & 5-th percentile of the {\sc SpectroTranslator} metallicity\\
   {\sc M\_H\_pred\_16} &   & 16-th percentile of the {\sc SpectroTranslator} metallicity\\
   {\sc M\_H\_pred\_50} &   & 50-th percentile of the {\sc SpectroTranslator} metallicity\\
   {\sc M\_H\_pred\_84} &   & 84-th percentile of the {\sc SpectroTranslator} metallicity\\
   {\sc M\_H\_pred\_95} &   & 95-th percentile of the {\sc SpectroTranslator} metallicity\\
   {\sc M\_H\_pred\_ERR} &   & Systematic error of the {\sc SpectroTranslator} on the metallicity\\
   & & \\
   {\sc MG\_FE\_pred\_5} &   & 5-th percentile of the {\sc SpectroTranslator} magnesium over iron ratio\\
   {\sc MG\_FE\_pred\_16} &   & 16-th percentile of the {\sc SpectroTranslator} magnesium over iron ratio\\
   {\sc MG\_FE\_pred\_50} &   & 50-th percentile of the {\sc SpectroTranslator} magnesium over iron ratio\\
   {\sc MG\_FE\_pred\_84} &   & 84-th percentile of the {\sc SpectroTranslator} magnesium over iron ratio\\
   {\sc MG\_FE\_pred\_95} &   & 95-th percentile of the {\sc SpectroTranslator} magnesium over iron ratio\\
   {\sc MG\_FE\_pred\_ERR} &   & Systematic error of the {\sc SpectroTranslator} on the magnesium over iron ratio\\
   & & \\
   {\sc VHELIO\_pred\_5} & km~s$^{-1}$  & 5-th percentile of the {\sc SpectroTranslator} heliocentric velocity\\
   {\sc VHELIO\_pred\_16} & km~s$^{-1}$  & 16-th percentile of the {\sc SpectroTranslator} heliocentric velocity\\
   {\sc VHELIO\_pred\_50} & km~s$^{-1}$  & 50-th percentile of the {\sc SpectroTranslator} heliocentric velocity\\
   {\sc VHELIO\_pred\_84} & km~s$^{-1}$  & 84-th percentile of the {\sc SpectroTranslator} heliocentric velocity\\
   {\sc VHELIO\_pred\_95} & km~s$^{-1}$  & 95-th percentile of the {\sc SpectroTranslator} heliocentric velocity\\
   {\sc VHELIO\_pred\_ERR} & km~s$^{-1}$  & Systematic error of the {\sc SpectroTranslator} on the heliocentric velocity\\
   & & \\
   {\sc Flag\_missing\_inputs\_intrinsic} & Boolean & If True, indicates that an input was missing for the {\it intrinsic} network\\
   {\sc Flag\_missing\_inputs\_extrinsic} & Boolean & If True, indicates that an input was missing for the {\it extrinsic} network\\
   {\sc Qflag\_Input\_intrinsic} & Boolean & Quality flag that indicates that the all the input of the {\it intrinsic} network are inside the parameter space covered by the training sample \\
   {\sc Qflag\_Output\_intrinsic} & Boolean & Quality flag that indicates that the all the Output of the {\it intrinsic} network are inside the parameter space covered by the training sample \\
   {\sc Qflag\_Input\_extrinsic} & Boolean & Quality flag that indicates that the all the input of the {\it extrinsic} network are inside the parameter space covered by the training sample \\
   {\sc Qflag\_Output\_extrinsic} & Boolean & Quality flag that indicates that the all the Output of the {\it extrinsic} network are inside the parameter space covered by the training sample \\
    {\sc Qflag\_comments} & String & Explication of why at least one of the quality flag was set on False\\
\hline
\end{tabular}
  \caption{Metadata for all sources from the GALAH DR3 catalogue transformed onto the APOGEE-~2 DR17 base.}
   \label{tab:descrip_cat}
\end{table}

\twocolumn
\section{Variation of the feature importance as function of the metallicity}\label{annex:featImp}

\begin{figure*}[h!]
\centering
  \includegraphics[angle=0,clip,width=16cm,viewport=80 0 935 430]{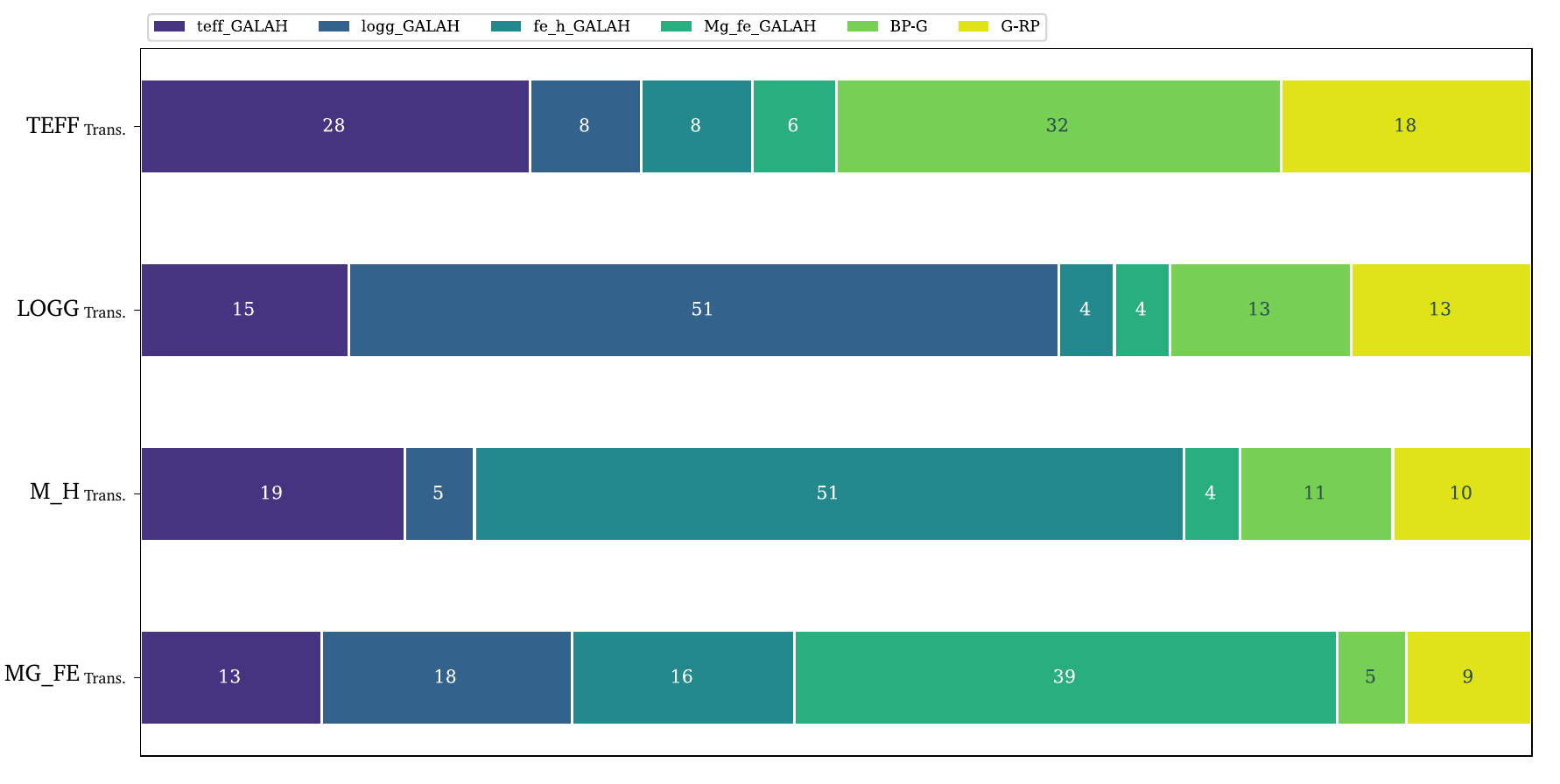}
   \caption{Same as Fig.~\ref{fig:shap_intrinsic} but for metal-poor stars ([Fe/H]$<-1.0)$.}
\label{fig:shap_metpoor}
\end{figure*}

\begin{figure*}[h]
\centering
  \includegraphics[angle=0,clip,width=16cm,viewport=80 0 935 430]{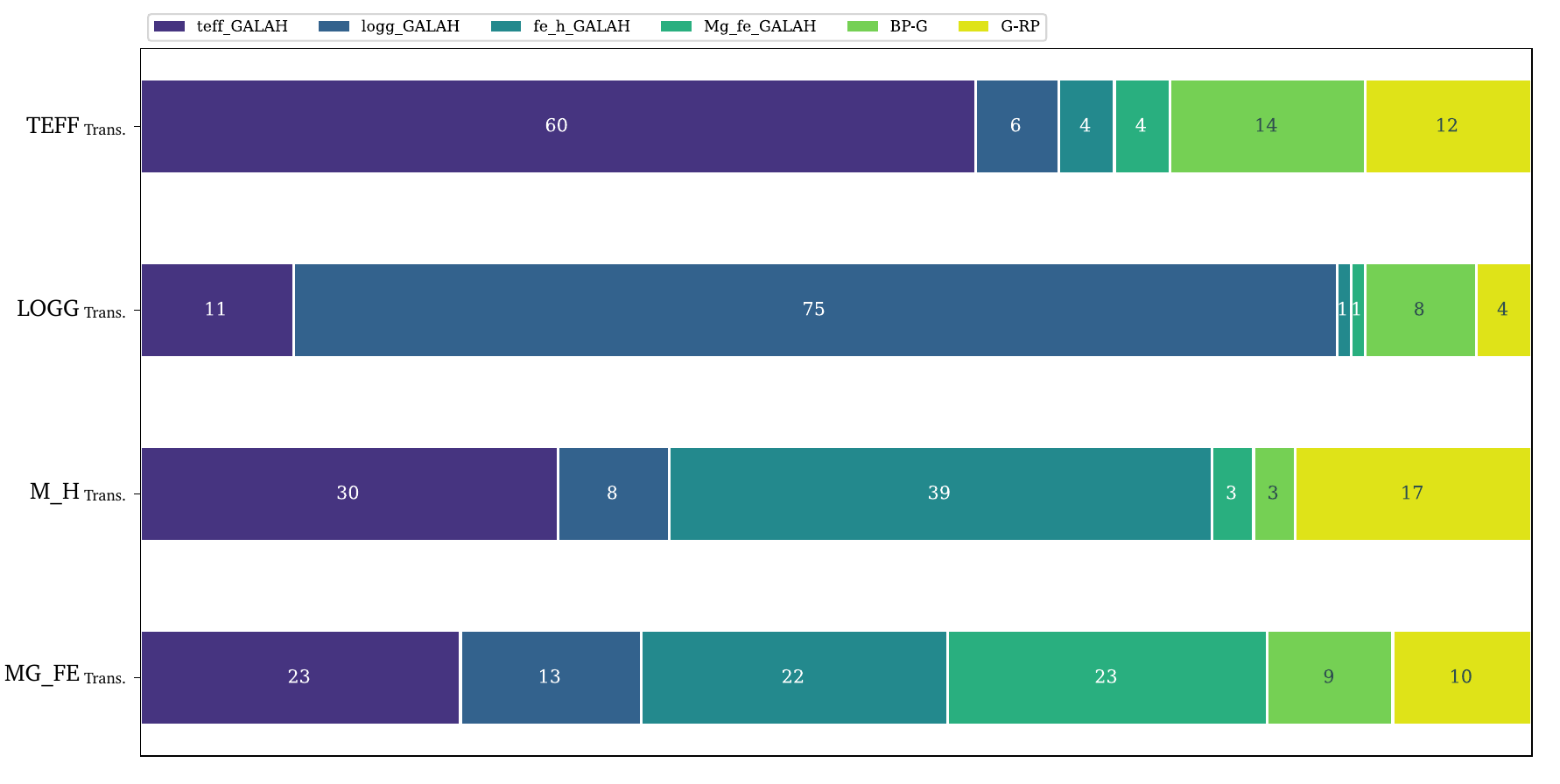}
   \caption{Same as Fig.~\ref{fig:shap_intrinsic} but for metal-rich stars ([Fe/H]$>-0.4)$.}
\label{fig:shap_metrich}
\end{figure*}

As discussed in Sect.~\ref{sec:shap}, the method used to estimate the importance of each input feature corresponds to their average importance throughout the entire parameter space, computed using a sample of 100 randomly selected stars from the validation set. In the case presented here, one can see that the parameter space is not homogeneously covered (e.g. Fig.~\ref{fig:comparison}), in particular for the metallicity, with $\simeq 80\%$ of the stars of the training/validation sample with $[$Fe/H$]>-0.4$. In case of such heterogeneous distribution, the mean absolute SHAP values correspond to the average importance of each parameter in the region of the parameter space the most populated in the validation sample. 

This is very clear when we compare Figs.~\ref{fig:shap_metpoor} and \ref{fig:shap_metrich}, where we show the mean absolute SHAP values computed using only metal-poor ($[$Fe/H$]<-1.0$) and metal-rich ($[$Fe/H$]>-0.4$) stars, with Fig.~\ref{fig:shap_intrinsic}, which was obtained using 100 randomly selected stars over the entire metallicity range. One can clearly see that the mean absolute SHAP values of the entire sample are very similar to the values obtained using only metal-rich stars. On the contrary, we see that in the metal-poor range these values are very different, with the transformation of the effective temperature from the GALAH to the APOGEE-~2 base that is more dependent on the photometric colours with respect to that of the metal-rich stars, and the transformation of the surface gravity that is more impacted by the colours, the metallicity, and [Mg/Fe]. For the transformation of the metallicity, the input metallicity plays a higher role for the metal-poor stars than for the metal-rich, although the importance of all the other parameters is relatively similar between the two regions. 

For the metal-poor stars, the original GALAH values of [Mg/Fe] play a more significant role in the transformation of [Mg/Fe] itself than for the metal-rich ones. This is intriguing, as one might have anticipated the opposite, with [Mg/Fe] being more influential for the metal-rich stars than for the metal-poor ones. This expectation arises from the clear bimodality observed in the high/low [Mg/Fe] distribution among the former, in contrast to the latter. This example underscores the complexity of interpreting these values. However, it is conceivable that this result is linked to the disparity in the parameter space covered by the metal-poor and metal-rich samples. Indeed, the metal-poor sample is mainly composed of giant stars spanning a narrow range of temperature, whereas the metal-rich sample includes both dwarf and giant stars covering a wider range of temperature.

As mentioned in Sect.~\ref{sec:shap}, the interpretability of neural networks is improving rapidly, and a significant improvement of the \textsc{SpectroTranslator} algorithm can be achieved through a better understanding of it and its interpretability.

\section{[Fe/H] and [Mg/Fe] distribution without the transformation of the {\sc SpectroTranslator}}

\begin{figure*}
\centering
  \includegraphics[angle=0,clip, viewport= 0 0 1050 480,width=18cm]{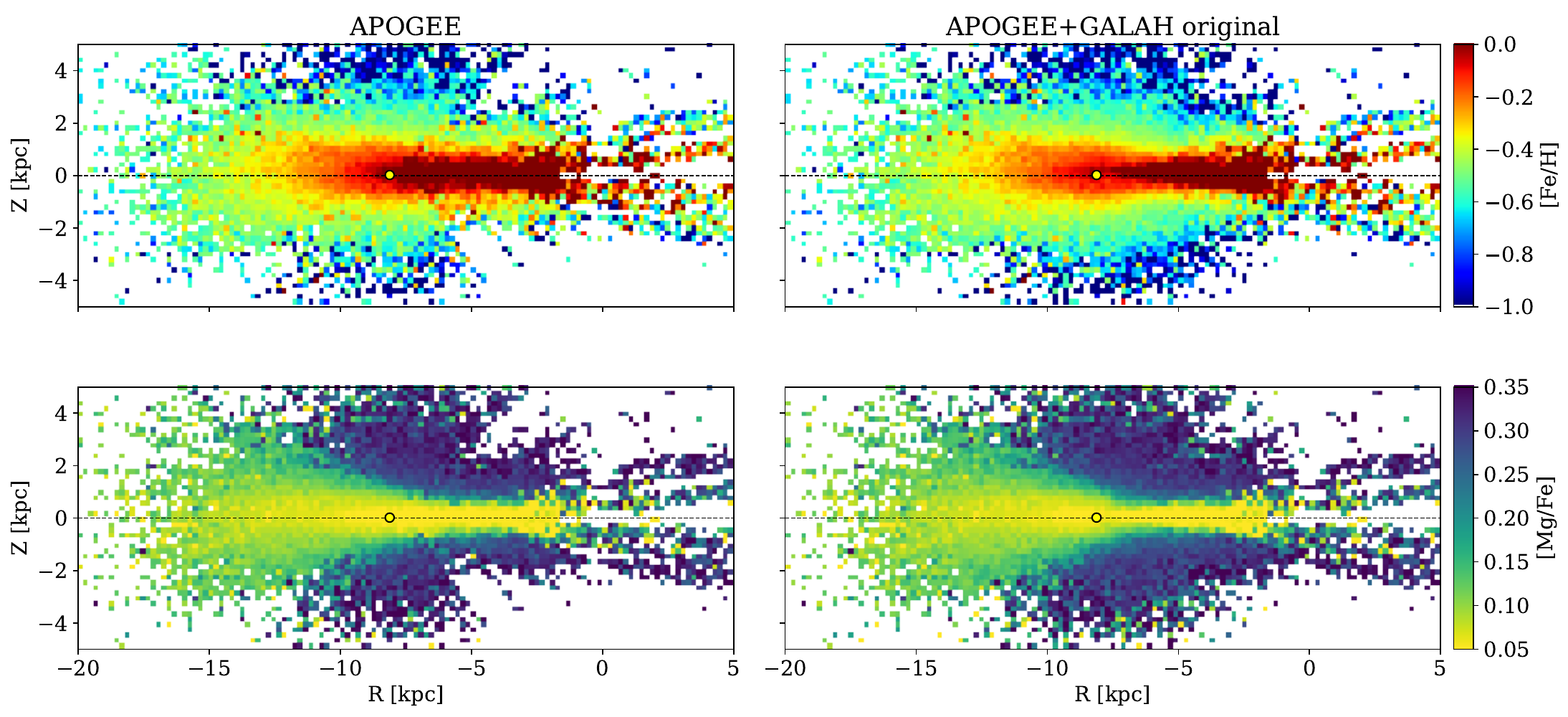}
   \caption{Same as Fig.~\ref{fig:Map_APOGEE_GALAH} but with the ``original'' [Fe/H] and [Mg/Fe] from GALAH.}
\label{fig:map_GALAHori}
\end{figure*}

In Fig.~\ref{fig:map_GALAHori}, we show the [Fe/H] and [Mg/Fe] distribution across the Galaxy by combining the APOGEE-~2 and GALAH samples, but using the ``original'' values for the latter, instead of the transformed values as in the Sect.~\ref{sec:gradient}.
Contrary to what shown in Sect.~\ref{sec:gradient}, one can see that without the transformation made by the {\sc SpectroTranslator} algorithm, the median [Fe/H] and [Mg/Fe] are very different when the APOGEE-~2 and GALAH data are combined than when the only APOGEE-~2 data are used. In particular, we can see that close to the Galactic midplane in the inner Galaxy, the median [Fe/H] and [Mg/Fe] values present a significant change, with a more conic profile when the GALAH data are used. Furthermore, we can see that the regions without APOGEE-~2 data stand apart to the other area. This is particularly visible in R$=-5$~kpc Z$=2$~kpc in both [Fe/H] and [Mg/Fe]. Comparing this figure with Fig.~\ref{fig:Map_APOGEE_GALAH} shows the strength of the {\sc SpectroTranslator} algorithm, since when the algorithm is used we preserved the global structure of the Galaxy measured with the APOGEE-~2 data only, contrary to when the ``original'' GALAH data are used. Nevertheless, it is interesting to note here that the vertical asymmetry in the median metallicity reported in Sect.~\ref{sec:gradient} is still visible even when using the ``original'' GALAH data, since most of the stars in the outer disc originate from the APOGEE-~2 sample.

\section{Radial and vertical gradients in APOGEE-~2 only} \label{sec:grad_APOGEE}

\begin{figure*}[!ht]
\centering
  \includegraphics[angle=0,clip,width=18.0cm]{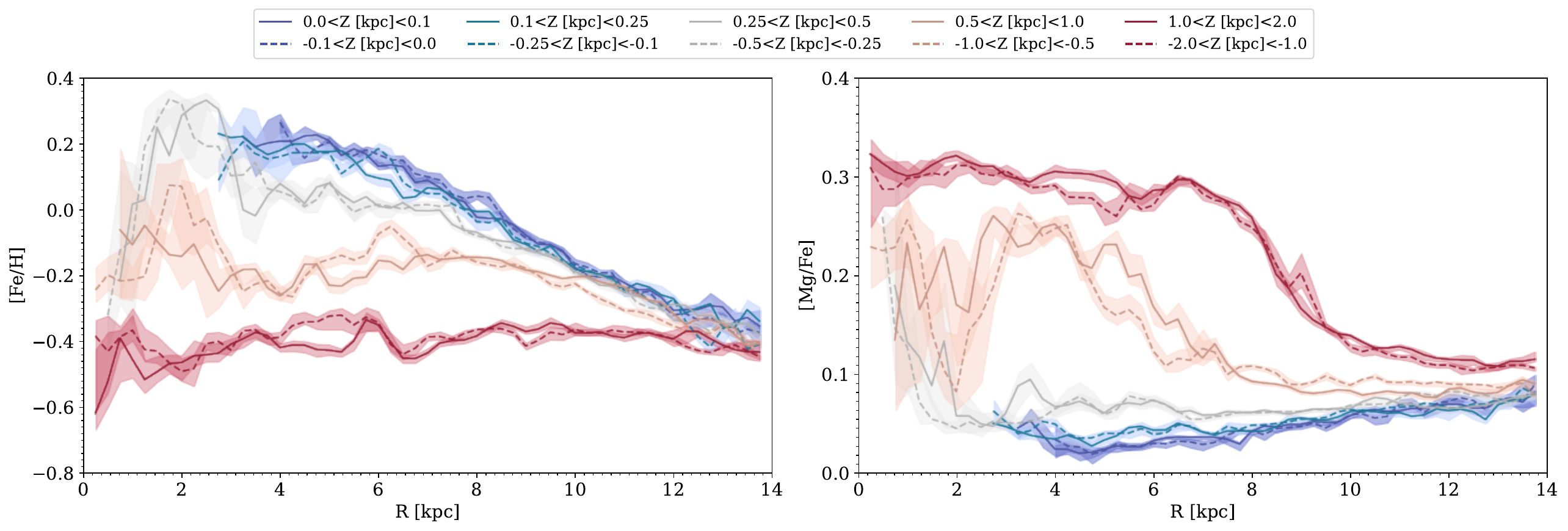}
   \caption{Same as Fig.~\ref{fig:MH_MG_R} but using only the stars from APOGEE-~2.}
\label{fig:MH_MG_R_APOGEE}
\end{figure*}

\begin{figure*}[!ht]
\centering
  \includegraphics[angle=0,clip,width=18.0cm]{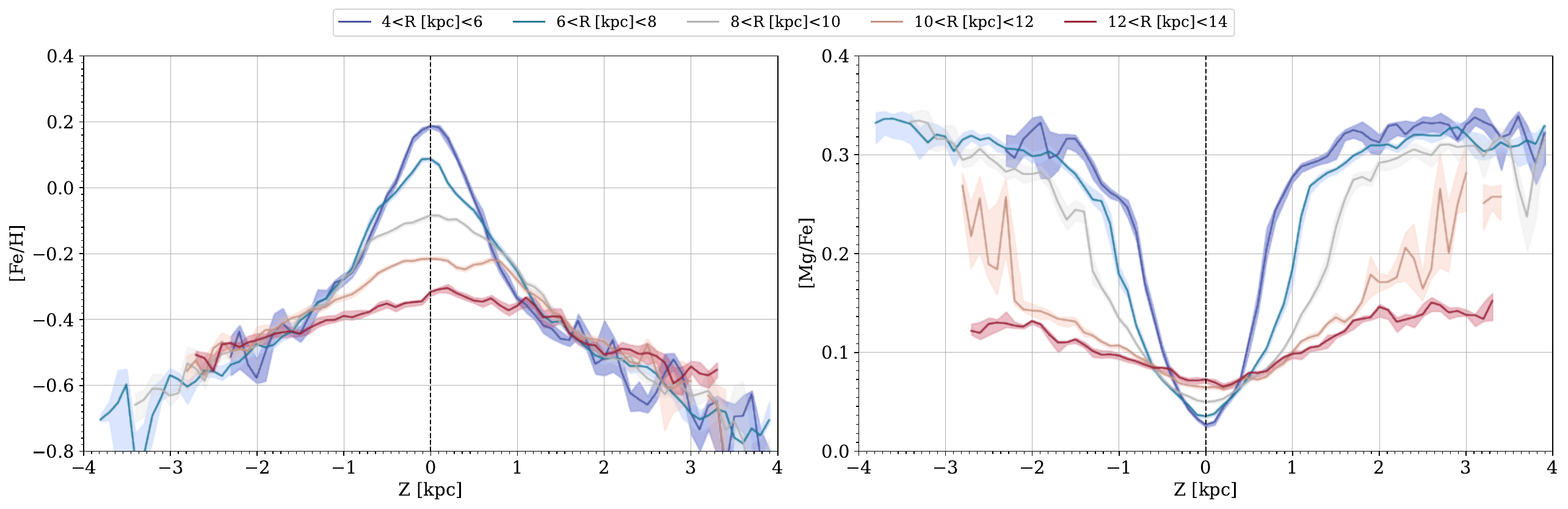}
   \caption{Same as Fig.~\ref{fig:MH_MG_Z} but using only the stars from APOGEE-~2.}
\label{fig:MH_MG_Z_APOGEE}
\end{figure*}

We reproduced here the analysis conducted in Sect.~\ref{sec:gradient}, but using only the stars observed by APOGEE-~2. Fig~\ref{fig:MH_MG_R_APOGEE}, and Fig.~\ref{fig:MH_MG_Z_APOGEE} show the radial and vertical gradients of the [Fe/H] and [Mg/Fe], respectively. Comparing them to Fig.~\ref{fig:MH_MG_R} and Fig.~\ref{fig:MH_MG_Z}, one can see that the global trends are very similar. However, a few differences are visible. In the radial median [Fe/H] and [Mg/Fe] gradient, we see that there is an asymmetry between the two sides of the disc between $4<$R$<6$~kpc for $1.0<|$Z$|<2.0$~kpc and between $5<$R$<7$~kpc for $0.5<|$Z$|<1.0$~kpc, which are less visible when we combined the APOGEE-~2 data with the transformed GALAH data. This difference might be due to statistical fluctuation, as there is a significant lower number of stars observed by APOGEE-~2 in the Southern Galactic hemisphere than in the North (see e.g. upper panels in Fig.~\ref{fig:Map_APOGEE_GALAH}), and adding the GALAH data reduce these fluctuations. This might also explain why the vertical median [Fe/H] between  $6<$R$<8$~kpc seems to present an asymmetry in the APOGEE-~2 data, and not when they are combined to the GALAH data. 

This comparison shows how the scientific legacy of a spectroscopic survey can be expanded by combining it with data obtained by another survey, after being calibrated on the same base.

\end{document}